\documentclass[12pt,draft]{article}

\usepackage{amsthm,amssymb,amsmath,amscd,amstext,mathrsfs}
\usepackage{latexsym,mathptmx,xypic,tikz}

\newtheorem{definition}{Definition}[section]
\newtheorem{theorem}{Theorem}[section]
\newtheorem{lemma}{Lemma}[section]
\newtheorem*{remark}{{\it Remark}}

\newcommand{\nc}{\newcommand} 

\nc{\C}{{\mathbb C}}
\nc{\R}{{\mathbb R}}
\nc{\HH}{{\mathbb H}}
\nc{\Z}{{\mathbb Z}}
\nc{\N}{{\mathbb N}}
\nc{\dd}{{\rm d}}

\nc{\ii}{{\bf i}}
\nc{\cg}{{\mathscr G}}

\begin{document}

\title{Exotica and the status of the strong cosmic censor conjecture 
in four dimensions}

\author{G\'abor Etesi\\
\small{{\it Department of Geometry, Mathematical Institute, Faculty of 
Science,}}\\
\small{{\it Budapest University of Technology and Economics,}}\\
\small{{\it Egry J. u. 1, H \'ep., H-1111 Budapest, Hungary}}
\footnote{e-mail: {\tt etesi@math.bme.hu}}}

\maketitle
                          
\pagestyle{myheadings}
\markright{Strong cosmic censorship in four dimensions and exotica}

\thispagestyle{empty}

\begin{abstract} An immense class of physical counterexamples 
to the four dimensional strong cosmic censor conjecture---in its usual 
broad formulation---is exhibited. More precisely, out of any closed and simply 
connected $4$-manifold an open Ricci-flat Lorentzian $4$-manifold is 
constructed which is not globally hyperbolic and no perturbation of it, in 
any sense, can be globally hyperbolic. This very stable 
non-global-hyperbolicity is the consequence of our open spaces 
having a ``creased end'' i.e., an end diffeomorphic to an exotic $\R^4$. 
Open manifolds having an end like this is a typical phenomenon in four 
dimensions.  

The construction is based on a collection of results of Gompf and Taubes 
on exotic and self-dual spaces, respectively, as well as applying 
Penrose' non-linear graviton construction (i.e., twistor theory) to 
solve the Riemannian Einstein's equation. These solutions then are 
converted into stably non-globally-hyperbolic Lorentzian vacuum 
solutions. It follows that the {\it plethora} of vacuum solutions we 
found cannot be obtained via the initial value formulation of the 
Einstein's equation because they are ``too long'' in a certain sense 
(explained in the text). This different (i.e., not based on the 
initial value formulation but twistorial) technical background 
might partially explain why the existence of vacuum solutions of this 
kind has not been realized so far in spite of the fact that, apparently, 
their superabundance compared to the well-known globally hyperbolic 
vacuum solutions is overwhelming. 
\end{abstract}

\centerline{AMS Classification: Primary: 83C75, Secondary: 57N13, 53C28}
\centerline{Keywords: {\it Strong cosmic censor conjecture; Exotic $\R^4$; 
Twistors}}


\section{Introduction}
\label{one}


Certainly one of the deepest open problems of contemporary classical 
general relativity is the validity or invalidity of the {\it strong 
cosmic censor conjecture} \cite{sim-pen}. This is not only a 
technical conjecture of a particular branch of current theoretical 
physics: it deals with the very foundations of our rational description 
of Nature. Indeed, Penrose' original aim in the 1960-70's with formulating this 
conjecture was to protect causality in generic gravitational situations. 
We have the strong conviction that in the {\it classical} physical world at 
least, every physical event (possibly except the 
initial Big Bang) has a physical cause which is another and preceding 
physical event. Since mathematically speaking space-times having this 
property are called {\it globally hyperbolic}, our requirement can be 
formulated roughly as follows (cf. e.g. \cite[p. 304]{wal}): 
\vspace{0.1in}

\noindent {\bf SCCC.} {\it A generic (i.e., stable), physically relevant
(i.e., obeying some energy condition) space-time is globally hyperbolic.}
\vspace{0.1in}

\noindent We do not make an attempt here to survey the vast physical and 
mathematical literature triggered by the {\bf SCCC} instead we refer to 
surveys \cite{ise, rin, pen3}. Rather we may summarize the current situation 
as follows. During the course of time the originally single {\bf SCCC} has 
fallen apart into several mathematical or physical 
versions, variants, formulations. For example there exists a generally 
working, mathematically meaningful but from a physical viewpoint rather 
weak version formulated in \cite[p. 305]{wal} and proved in \cite{ete2}. 
In another approach to the {\bf SCCC} based on initial value formulation 
\cite[Chapter 10]{wal}, on the one hand, there are certain specific classes of 
space-times in which the {\bf SCCC} allows a mathematically rigorous as well 
as physically contentful formulation whose validity can be established 
\cite{rin}; on the other hand counterexamples to the {\bf SCCC} in this 
formulation also regularily appear in the literature however they are 
apparently too special, not ``generic''. In spite of these sporadic 
counterexamples the overall confidence in the physicist community is 
that an appropriate form of the {\bf SCCC} must hold true hence causality is 
saved.

However here we claim to exhibit an {\it abundance of generic counterexamples 
to the {\bf SCCC}} whose first agent was announced in \cite{ete1}. 
Informally speaking, the content of our main results here, namely Theorems 
\ref{riemannverzio} and \ref{lorentzverzio} can be summarized as follows:
\vspace{0.1in}

\noindent $\overline{\bf SCCC}${\bf .} {\it From every connected and simply
connected closed (i.e., compact without boundary) smooth $4$-manifold $M$
one can construct an open (i.e., non-compact without boundary) smooth
$4$-manifold $X_M$ and a smooth Ricci-flat 
Lorentzian metric $g$ on it such that $(X_M,g)$ is not globally hyperbolic.
Moreover, any ``sufficiently large'' (in an appropriate topological sense)
physical perturbation $(X'_M,g')$ of this space cannot be globally
hyperbolic, too.

This very stable non-global-hyperbolicity follows because $X_M$
as a smooth $4$-manifold contains a ``creased end'' (see Figure 1), a
typical four dimensional phenomenon.}
\vspace{0.1in}

\noindent What is then the resolution of the apparent contradiction 
between the well-known affirmative solutions and our negative result 
$\overline{\bf SCCC}$ here? In this short introduction we just would like to 
draw attention to a historical aspect of the answer and try to offer more 
technical comments at the end of the paper. The Einstein equation as a 
non-linear partial differential equation on a $4$-manifold is a quite 
transcendental object in the sense that there is yet no systematic way 
to solve it. So far the initial value formulation is the only known method 
which can provide sufficiently many solutions in various situations hence 
its investigation by Leray, Choquet-Bruhat, Lichnerowicz, Geroch in the 
1950-60's and by many others later cannot be overestimated. The initial value 
formulation starts off by considering an initial data set $(S,h,k)$ with 
$S$ being a smooth {\it three dimensional} manifold and $h,k$ certain tensor 
fields on it satisfying (simpler) constraint equations; and out of these 
data it produces a solution $(M,g)$ of the Lorentzian Einstein equation. An 
apparently innocuous technical by-product of the initial value formulation 
is that it fixes not only the metric but the smooth structure 
of the resulting space-time, too: the underlying 
{\it four dimensional} manifold $M$ is always {\it diffeomorphic} to the 
product $S\times\R$ (with their unique smooth structures) by the celebrated 
Bernal--S\'anchez theorem \cite{ber-san}. This technical nuance seemed to 
be not a problem at all for the physicist community by the time 
the initial value formulation came to existence. 

Side-by-side with but quite isolated from these investigations 
mathematicians also made efforts to understand the structure of smooth 
manifolds and they came up with unexpected issues. Since the early works of 
Whitney, Milnor in the 1950-60's followed by Casson, Kirby and others, it 
had been gradually realized that in higher dimensions topology and 
smoothness do not determine each other and their interaction gets particularly 
complicated in four dimensions. By the early 1980's it was recognized that 
essentially no known compact smoothable topological $4$-manifold carries 
exactly one smooth structure; in fact in most of the well-understood cases 
they admit not only more than one but countably infinitely many different ones 
\cite{gom-sti}. In the case of non-compact (relevant for physics) topological 
$4$-manifolds there is even no obstruction against smooth structure and they 
typically accommodate an uncountably family of them \cite{gom3}. A 
characteristic feature of these ``exotic'' smooth 
structures---unforeseenable in the 1950's---is that 
they are ``creased'' i.e., do not arise as smooth products of lower 
dimensional smooth structures. The striking discovery of 
exotic (or fake) $\R^4$'s (i.e., smooth $4$-manifolds which are homeomorphic 
but not diffeomorphic to the usual $\R^4$) by Donaldson, Freedman, Taubes in 
the 1980's and investigated further by Akbulut, Bi\v{z}aca, Gompf and 
others during the 1990's and 2000's is probaly the most dramatic example 
of the general situation completely absent in other dimensions.

From this perspective the cases for which the validity of the ${\bf SCCC}$ 
has been verified so far \cite{ise, rin} seem to be atypical hence 
essentially negligable ones; partly because affirmative answers have been 
obtained by the initial value formulation hence the underlying 
space-times in these affirmative solutions are never ``creased''. On the 
contrary, our counterexample factory $\overline{\bf SCCC}$ rests on 
typical features of smooth $4$-manifolds. The only way to refute the 
general position adopted here when dealing with the ${\bf SCCC}$ is if one 
could somehow argue that general smooth $4$-manifolds are too ``exotic'', 
``fake'' or ``weird'' from the aspect of physical general relativity. 
However from the physical viewpoint if the ``summing over everything'' 
approach to quantum gravity is correct then very general unconventional but 
still physical space-times should be considered, too \cite{ass,dus}; from the 
mathematical perspective non-linear partial differential equations like 
Einstein's equation are typically also solvable. Consequently both 
physically and mathematically speaking the true properties of general 
relativity cannot be revealed by understanding it only on simple atypical 
manifolds; the division of smooth $4$-manifolds into ``usual'' and 
``unusual'' ones can be justified only by conventionalism i.e., one has to 
evoke historical (and technical) arguments to pick up ``usual'' spaces from 
the bottomless sea of smooth $4$-manifolds and abandon others. In fact 
our general position supporting $\overline{\bf SCCC}$ fits well 
with the {\it four dimensional} (i.e., the original Einstein--Hilbert) 
{\it Lagrangian} formulation while the initial value formulation supporting 
{\bf SCCC} rests on a {\it three dimensional Hamiltonian} reformulation of 
general relativity. Therefore, to summarize, in our opinion the choice between 
{\bf SCCC} or $\overline{\bf SCCC}$ reflects one's commitment toward one 
of these formulations of general relativity. 

This paper, considered as a substantial generalization and technical 
improvement of \cite{ete1}, is organized as follows. Section \ref{two} offers 
very general definitions of what one would expect to mean by a perturbation 
of a space-time and a counterexample to the {\bf SCCC}. Section \ref{three} 
contains the construction of complete Ricci-flat {\it Riemannian} manifolds 
based on twistor theory. Section \ref{four} describes how to convert 
these solutions into stably non-globally-hyperbolic Ricci-flat 
{\it Lorentzian} manifolds i.e., counterexamples. Section \ref{five} 
contains some concluding remarks and an outlok. Section \ref{six} is an 
Appendix with a summary of the theory of Lebesgue integration in algebraic 
function fields, a tool has been used in Section \ref{three}. 

Our notational convention throughout the text is that $\R^4$ will denote 
the four dimensional real vector space equipped with its standard 
differentiable manifold structure whilst $R^4$ or $R^4_t$ will denote 
various exotic (or fake) variants. The notation ``$\:\cong\:$'' will 
always mean ``diffeomorphic to'' whilst homeomorphism always will be 
spelled out as ``homeomorphic to''. Finally we note that all set 
theoretical or topological operations (i.e., $\subseteqq$, $\cap$, $\cup$, 
taking open or closed subsets, closures, complements, etc.) will be taken 
in a manifold $M$ with its well-defined standard manifold topology throughout 
the text. In particular for $\R^4$ or the $R^4_t$'s this topology is the 
unique underlying manifold topology.


\section{Definition of a counterexample}
\label{two}


In agreement with the common belief in the physicist and mathematician 
community, formulating the strong cosmic censor conjecture in a 
mathematically rigorous way is obstructed by lacking an overall 
satisfactory concept of ``genericity''. Consequently the 
main difficulty to find a ``generic counterexample'' to the {\bf SCCC} lies 
not in its actual finding (indeed, most of the well-known basic 
solutions of Einstein's equation provide violations of it) but rather in 
proving that the particular counterexample is ``generic''. In this section 
we outnavigate this problem by mathematically formulating the concept 
of a certain counterexample which is logically stronger than a ``generic 
counterexample'' to the {\bf SCCC}. Then we search for a counterexample of 
this kind making use of uncountably many large exotic $\R^4$'s. 

A standard reference here is \cite[Chapters 8,10]{wal}. By a {\it space-time} 
we mean a connected, four dimensional, smooth, time-oriented 
Lorentzian manifold without boundary. By a {\it (continuous) Lorentzian 
manifold} we mean the same thing except that the metric is allowed to be a 
continuous tensor field only.

\begin{definition} Let $(S, h, k)$ be an initial data set for Einstein's 
equation with $(S, h)$ a connected complete Riemannian $3$-manifold and with a 
fundamental matter represented by a stress-energy tensor $T$ obeying the 
dominant energy condition. Let $(D(S), g\vert_{D(S)})$ be the unique maximal 
Cauchy development of this initial data set. Let $(M,g)$ be a further maximal 
extension of $(D(S), g\vert_{D(S)})$ as a (continuous) Lorentzian manifold if 
exists. That is, $(D(S), g\vert_{D(S)})\subseteqq (M,g)$ is a (continuous) 
isometric embedding which is proper if $(D(S), g\vert_{D(S)})$ is 
still extendible and $(M,g)$ does not admit any further proper isometric 
embedding. (If the maximal Cauchy development is inextendible then 
put simply $(M,g):=(D(S), g\vert_{D(S)})$ for definiteness.)

The (continuous) Lorentzian manifold $(M',g')$ is a {\rm perturbation of 
$(M,g)$ relative to $(S,h,k)$} if 
\begin{itemize}

\item[{\rm (i)}] $M'$ has the structure 
\[M':=\mbox{{\rm the connected component of $M\setminus H$ containing $S$}}\]
where, for a connected open subset $U$ of $M$ containing the initial surface 
i.e., $S\subset U\subseteqq M$, the subset $H$ is closed and satisfies 
$\emptyset\subseteqq H\subseteqq\partial U=\overline{U}\setminus U$ i.e., is 
a closed subset in the boundary of $U$ (consequently $M'\subseteqq M$ is open 
hence inherits a differentiable manifold structure);

\item[{\rm (ii)}] $g'$ is a solution of Einstein's 
equation at least in a neighbourhood of the initial surface 
$S\subset M'$ with a fundamental matter represented by a stress-energy tensor 
$T'$ obeying the dominant energy condition at least in a neighbourhood of 
$S\subset M'$; 

\item[{\rm (iii)}] $(M',g')$ does not admit further proper isometric embeddings 
and $(S,h')\subset (M',g')$ with $h':=g'\vert_S$ is a spacelike complete 
sub-$3$-manifold. 
\end{itemize}

\label{perturbacio}
\end{definition}

\begin{remark}\rm 1. It is crucial that in the original spirit of relativity 
theory we consider metric perturbations of the {\it four} dimensional 
space-time $M$ (whilst keeping its underlying smooth structure fixed)---and 
not those of a {\it three} dimensional initial data set. This natural class 
of perturbations is therefore immense: it contains all connected manifolds 
$M'$ satisfying 
\[S\subset M'\subseteqq M\] 
i.e., contain the initial surface but perhaps being topologically different 
from the original manifold. The perturbed metric is a physically relevant 
solution of Einstein's equation at least in the vicinity of $S\subset M'$ 
such that $(M',g')$ is inextendible and $(S,h')\subset (M',g')$ is still 
spacelike and complete. In other words these perturbations are physical 
solutions allowed to blow up along certain closed `` boundary subsets'' 
$\emptyset\subseteqq H\subset M$; the notation $H$ for these subsets 
indicates that among them the (closure of the) Cauchy horizon $H(S)$ of 
$(S,h,k)$ may also appear. Beyond the non-singular perturbations satisfying 
$H=\emptyset$ a prototypical example with 
$H\not=\emptyset$ is the physical perturbation $(M',g')$ of the 
(maximally extended) undercharged Reissner--Nordstr\"om space-time $(M,g)$ 
by taking into account the full backreaction of a pointlike particle or any 
classical field put onto the originally pure electro-vacuum space-time 
(``mass inflation''). In this case the singularity subset $H$ is expected to 
coincide with the (closure of the) full inner event horizon of the 
Reissner--Nordstr\"om black hole which is the Cauchy horizon for the standard 
initial data set inside the maximally extended space-time \cite{sim-pen}. 
A similar perturbation of the Kerr--Newman space-time is another example 
with $H\not=\emptyset$.

2. Accordingly, note that in the above definition of perturbation none 
of the terms ``generic'' or ``small'' have been used. This indicates 
that if such types of perturbations can be somehow specified then one 
should be able to recognize them among the very general but still physical 
perturbations of a space-time as formulated in Definition \ref{perturbacio}. 
\end{remark}

\noindent Now we are in a position to formulate in a mathematically 
precise way what we mean by a ``robust counterexample'' to the {\bf SCCC} 
as formulated roughly in the Introduction. 

\begin{definition} Let $(S, h, k)$ be an initial data set for Einstein's 
equation with $(S, h)$ a connected complete Riemannian $3$-manifold and with 
a fundamental matter represented by a stress-energy tensor $T$ obeying the 
dominant energy condition. Assume that the maximal Cauchy development of this 
initial data set is extendible i.e., admits a (continuous) isometric 
embedding as a proper open submanifold into an inextendible (continuous) 
Lorentzian manifold $(M,g)$.

Then $(M,g)$ is a {\rm robust counterexample to the {\bf SCCC}} 
if it is very stably non-globally hyperbolic i.e., all of its perturbations 
$(M',g')$ relative to $(S,h,k)$ are not globally hyperbolic.

\label{ellenpelda}
\end{definition}

\begin{remark}\rm 1. Concerning its logical status it is reasonable to 
consider this as a {\it generic} counterexample because the perturbation 
class of Definition \ref{perturbacio} is expected to contain all ``generic 
perturbations'' whatever they are. Consequently in Definition \ref{ellenpelda} 
we are dealing with a stronger statement than the logical negation of the 
affirmative sentence in {\bf SCCC}. 

2. The trivial perturbation i.e., the extension $(M,g)$ itself in 
Definition \ref{ellenpelda} cannot be globally hyperbolic as observed already 
in \cite[Remark after Theorem 2.1]{ete2}. 
\end{remark}


\section{Riemannian considerations}
\label{three}


Strongly influenced by \cite{ber-san, che-nem}, in order to attack the 
{\bf SCCC} we begin now our excursion 
into the weird world of the four dimensional exotic {\it m\'enagerie} (or 
rather {\it plethora}). A standard reference here is 
\cite[Chapter 9]{gom-sti}. Our aim in this section is to prove the following

\begin{theorem}
Let $M$ be a connected and simply connected, closed 
(i.e., compact without boundary) smooth $4$-manifold. Then out of this 
manifold one can construct a Riemannian $4$-manifold 
\[(X_M,g_0)\]
with the following properties. 

The metric $g_0$ is a smooth Ricci-flat 
complete Riemannian metric on $X_M$. Furthermore, the space $X_M$ is 
an open (i.e., non-compact without boundary) oriented smooth $4$-manifold 
with a single so-called creased end. Here ``creased'' means that if 
$S\subset X_M$ is an arbitrary smoothly embedded sub-$3$-manifold then 
$X_M\not\cong S\times\R$ i.e., $X_M$ does not split smoothly into the 
product of a $3$-manifold and $\R$ (with their unique smooth structures). 
\label{riemannverzio}
\end{theorem}
\noindent The proof of this theorem is based on a collection of strong 
and surprising results of Gompf \cite{gom1, gom2, gom3} and Taubes 
\cite{tau1,tau2} and is rather involved. Therefore our plan to prove it is as 
follows: first we recall these results in precise forms in order to 
have a solid starting point and then through a sequence 
of technical lemmata we will arrive at the proof of Theorem 
\ref{riemannverzio} at the end of this section. 

It is well-known that the Fubini--Study metric on the complex projective 
space $\C P^2$ with orientation inherited from its complex structure is 
self-dual (or half-conformally flat) i.e., the anti-self-dual part of its 
Weyl tensor vanishes; consequently the oppositely oriented complex projective 
plane $\overline{\C P^2}$ is anti-self-dual. A powerful generalization of 
this latter classical fact is Taubes' construction of an abundance of 
anti-self-dual $4$-manifolds; firstly we exhibit his result but now in an 
orientation-reversed form:

\begin{theorem} {\rm (Taubes \cite[Theorem 1.1]{tau2})} Let $M$ be a 
connected, compact, oriented smooth 
$4$-manifold. Let $\C P^2$ denote the complex projective plane with its 
usual orientation and let $\#$ denote the operation of taking the connected 
sum of manifolds. Then there exists a natural number $k_M\geqq 0$ such 
that for all $k\geqq k_M$ the modified compact manifold 
\[M\#\underbrace{\C P^2\#\dots\#\C P^2}_{k}\] 
admits a self-dual Riemannian metric. $\Diamond$

\label{taubestetel}
\end{theorem}

\noindent Secondly we evoke a result which is a sort of summary of what is so 
special in four dimensions (i.e., absent in any other ones): we recall a 
special class of large exotic (or fake) $\R^4$'s whose 
properties we will need here are summarized as follows:

\begin{theorem} {\rm (Gompf--Taubes, cf. \cite[Lemma 9.4.2, Addendum 9.4.4 and 
Theorem 9.4.10]{gom-sti})} There exists a pair $(R^4,K)$ consisting of a 
differentiable $4$-manifold $R^4$ homeomorphic but not diffeomorphic to the 
standard $\R^4$ and a compact oriented smooth $4$-manifold $K\subset R^4$ 
such that 
\begin{itemize}

\item[{\rm (i)}] $R^4$ cannot be smoothly embedded into the standard 
$\R^4$ i.e., $R^4\not\subseteqq\R^4$ but it can be smoothly embedded as a 
proper open subset into the complex projective plane i.e., 
$R^4\subsetneqq\C P^2$;

\item[{\rm (ii)}] Take a homeomorphism $f:\R^4\rightarrow R^4$, let 
$0\in B^4_t\subset\R^4$ be the standard open $4$-ball of radius $t\in\R^+$ 
centered at the origin and put $R^4_t:=f(B^4_t)$ and 
$R^4_{+\infty}:=R^4$. Then 
\[\left\{ R^4_t\:\left\vert\:\mbox{$r\leqq t\leqq +\infty$ such that 
$0<r<+\infty$ satisfies $K\subset R^4_r$}\right.\right\}\]
is an uncountable family of nondiffeomorphic exotic $\R^4$'s none of them 
admitting a smooth embedding into $\R^4$ i.e., $R^4_t\not\subseteqq\R^4$ 
for all $r\leqq t\leqq +\infty$. 

\end{itemize}

This class of manifolds is called the {\rm Gompf--Taubes large radial 
family}. $\Diamond$

\label{egzotikusnagycsalad}
\end{theorem}

\begin{remark}\rm 1. The fact that any member $R^4_t$ in this family is not 
diffeomorphic to $\R^4$ implies the counterintuitive phenomenon that 
$R^4_t\not\cong W\times\R$ i.e., $R^4_t$ does not admit any {\it smooth} 
splitting into a $3$-manifold $W$ and $\R$ (with their unique smooth 
structures) in spite of the fact that such {\it continuous} splittings 
obviously exist. Indeed, from the contractibility of $R^4_t$ we can see that 
$W$ must be a contractible open $3$-manifold (a so-called 
{\it Whitehead continuum}) however, by an early result of McMillen \cite{mcm}, 
spaces of this kind always satisfy $W\times\R\cong\R^4$ i.e., their product 
with a line is always diffeomorphic to the standard $\R^4$. We will call this 
property of (any) exotic $\R^4$ below as ``creased''. The existence of 
many non-homeomorphic Whitehead continua have interesting 
consequences in the initial value formulation, too cf. e.g. \cite{new-cla}.

2. From Theorem \ref{egzotikusnagycsalad} we deduce that 
for all $r<t<+\infty$ there is a sequence of smooth proper embeddings 
\[R^4_r\subsetneqq R^4_t\subsetneqq R^4_{+\infty}=R^4\subsetneqq\C P^2\] 
which are very wild in the following sense. The complement $\C P^2
\setminus R^4$ of the largest member $R^4$ of this family is 
homeomorphic to $S^2$ regarded as an only continuously embedded projective 
line in $\C P^2$; therefore, if not otherwise stated later, 
let us denote this complement as $S^2:=\C P^2\setminus R^4\subset\C P^2$ 
in order to distinguish it from the ordinary projective lines 
$\C P^1\subset\C P^2$. If $\C P^2=\C^2\cup\C P^1$ is any 
holomorphic decomposition then $R^4\cap\C P^1\not=\emptyset$ (because 
otherwise $R^4\subseteqq\C^2\cong\R^4$ would hold, a contradiction) as 
well as $S^2\cap \C P^1\not=\emptyset$ (because otherwise 
$H_2(R^4;\Z )\cong\Z$ would hold 
since $\C P^1\subset\C P^2$ represents a generator of $H_2(\C 
P^2;\Z)\cong\Z$). Hence any ordinary projective line in $\C P^2$ is 
intersected by both $R^4$ and its complementum $S^2$ in $\C P^2$. 
This demonstrates that the members of the large radial family ``live 
somewhere between'' $\C^2$ and its projective closure $\C P^2$. 
However a more precise identification or location of them is a 
difficult task because these large exotic $\R^4$'s---although being 
honest differentiable $4$-manifolds---are very transcendental objects 
\cite[p. 366]{gom-sti}: they require infinitely many $3$-handles in any 
handle decomoposition (like any other known large exotic $\R^4$) and 
there is presently\footnote{More precisely in the year 1999, cf. 
\cite{gom-sti}.} no clue as how one might draw explicit handle diagrams 
of them (even after removing their $3$-handles). We note that the 
structure of small exotic $\R^4$'s i.e., which admit smooth embeddings 
into $\R^4$, is better understood, cf. \cite[Chapter 9]{gom-sti}. 
\end{remark}

\noindent Our last ingredient is the following {\it m\'enagerie} 
result of Gompf.

\begin{theorem} {\rm (Gompf \cite[Theorem 2.1]{gom3})} Let $X$ be a 
connected (possibly non-compact, possibly with boundary) topological
$4$-manifold and let $X':=X\setminus\{\mbox{ one point of $X$}\}$ be 
the punctured manifold with a single point removed. Then the non-compact
space $X'$ admits noncountable many (with the cardinality of the
continuum in ZFC set theory) pairwise non-diffeomorphic smooth
structures. $\Diamond$

\label{gompftetel}
\end{theorem}

\begin{remark}\rm If for instance $X$ is a connected compact smooth 
$4$-manifold in Theorem \ref{gompftetel} then Gompf's construction goes 
as follows: take the maximal large $R^4$ from Theorem 
\ref{egzotikusnagycsalad} and put $X':=X\#R^4$. This smooth 
$4$-manifold is obviously homeomorphic to the punctured $X'$; more 
generally, $X'_t:=X\#R^4_t$ will produce uncountable many smooth structures 
on the unique topological $4$-manifold underlying $X'_t$. 
\end{remark}

\noindent For our purposes and to begin with, we combine Theorems 
\ref{taubestetel}, \ref{egzotikusnagycsalad} and \ref{gompftetel} together as 
follows. 
\begin{lemma}

Out of any connected, closed (i.e., compact without boundary) oriented 
smooth $4$-manifold $M$ one can construct a connected, open (i.e., 
non-compact without boundary) oriented smooth Riemannian $4$-manifold 
$(X_M,g_1)$ which is self-dual but incomplete in general. 

Moreover $X_M$ has a single creased end where ``creased'' means that if 
$S\subset X_M$ is any smoothly embedded sub-$3$-manifold then 
$X_M\not\cong S\times\R$ i.e., $X_M$ does not split 
smoothly into the product of any smooth $3$-manifold $S$ and $\R$ (with 
their unique smooth structures).
\label{kezdolemma}
\end{lemma}

\noindent {\it Proof.} Pick any connected, oriented, closed, smooth 
$4$-manifold $M$. Referring to Theorem \ref{taubestetel} let $k:= \max 
(1,k_M)\in\N$ be a positive integer, put  
\[\hat{X}_M:=M\#\underbrace{\C P^2\#\dots\#\C P^2}_{k}\]
and let $\hat{g}_1$ be a self-dual metric on it. Then 
$(\hat{X}_M,\hat{g}_1)$ is a compact self-dual manifold. If 
$S^2=\C P^2\setminus R^4$ denotes the complement of $R^4\subset\C P^2$ as in 
the {\it Remark} after Theorem \ref{egzotikusnagycsalad} and $K\subset R^4$ is 
the compact subset as in part (ii) of Theorem \ref{egzotikusnagycsalad} then 
put 
\begin{equation}
X_M:=M\#\underbrace{\C P^2\#\dots\#\C P^2}_{k-1}\#_K(\C P^2\setminus S^2)\cong 
M\#\underbrace{\C P^2\#\dots\#\C P^2}_{k-1}\#_KR^4
\label{ujsokasag}
\end{equation}
where the operation $\#_K$ means that the attaching point $y_0\in R^4$ 
used to glue $R^4$ with $M\#\C P^2\#\dots\#\C P^2$ satisfies 
$y_0\in K\subset R^4$. The result is a connected, open 
$4$-manifold (see Figure 1). From the proper smooth embedding 
$X_M\cong \hat{X}_M\setminus S^2\subsetneqq\hat{X}_M$ there exists a 
self-dual Riemannian metric $g_1:=\hat{g}_1\vert_{X_M}$ on $X_M$ which is 
however in general non-complete. 

Although being non-compact, if $S\subset X_M$ is any smoothly 
embedded sub-$3$-manifold then obviously $X_M\not\cong S\times\R$ i.e., 
$X_M$ does not split {\it smoothly} into the product of a smooth 
$3$-manifold $S$ and $\R$ (with their standard smooth structures) due to 
its exotic $\R^4$-end i.e., the $R^4$-factor present in its 
decomposition (\ref{ujsokasag}) above. $\Diamond$ 
\vspace{0.3in}

\centerline{
\begin{tikzpicture}[scale=0.7]
\node at (-0.8,0) {$M$};
\draw [thick] plot [tension=0.8,smooth cycle] coordinates {
(0,0) (2,2) (5,0) (2,-2)};
\draw [thick] plot [tension=0.8,smooth] coordinates {(0.5,0.4) (1.5,0.2) 
(2.5,0.4)};
\draw [thick] plot [tension=0.8,smooth] coordinates {(0.8,0.3) (1.5,0.5) 
(2.2,0.3)};
\node at (7.2,0) {$X_M$};
\draw [thick] plot [tension=0.8,smooth] coordinates { (8.5,0.2) (9,-0.2) 
(9.5,0.2)};
\draw [thick] plot [tension=0.8,smooth] coordinates { (8.7,0) (9,0.1) 
(9.3,0)};
\draw [thick] plot [tension=0.8,smooth] coordinates 
{(8.5,1) (9,1.3) (10, 1) (11,0.7) 
(12,1) (12.5,1.5) (12,2) (13,2.5) (14,2) (13.1,1.3) (13.4,0.6) (14,0.3)};
\draw [thick] plot [tension=0.8,smooth] coordinates 
{(8.5,-1) (9,-1.3) (10, -1) (11,-0.7)
(12,-1) (12.5,-1.5) (12,-2) (13,-2.5) (14,-2) (13.1,-1.3) (13.4,-0.6) 
(14,-0.3)};
\draw [thick] plot [tension=0.8,smooth] coordinates {(8.5,1) (8,0) 
(8.5,-1)};
\draw [thick] plot [tension=0.8,smooth] coordinates {(14,0.3) (15,0.2) (16, 1)};
\draw [thick] plot [tension=0.8,smooth] coordinates {(14,-0.3) (15,-0.2) 
(16, -1)};
\draw[gray] plot coordinates {(16,1) (16.5,0.9) (16,0.7) (16.5,0.6) 
(16,0.5) (16.5,0.4) (16,0.3) (16.5,0.2) (16,0.1) (16.5,0) (16,-0.1) 
(16.5,-0.2) (16,-0.3) (16.5,-0.4) (16,-0.5) (16.5,-0.6) (16,-0.7) 
(16.5,-0.9) (16,-1)};
\end{tikzpicture}
}
\vspace{0.1in}

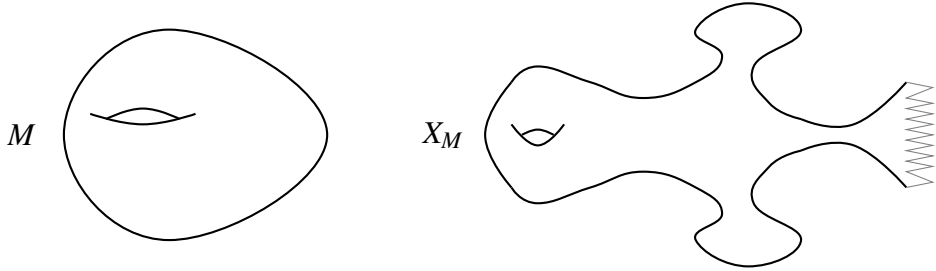
\centerline{Figure 1. Construction of $X_M$ out of $M$. 
The creased end of $X_M$ is drawn by a gray zig-zag.} 
\vspace{0.3in}

\noindent Next we improve the incomplete 
self-dual space $(X_M,g_1)$ of Lemma \ref{kezdolemma} to a complete 
Ricci-flat space $(X_M,g_0)$ by conformally rescaling $g_0$ with a 
suitable positive smooth function $\varphi :X_M\rightarrow\R^+$ which is a 
``multi-task'' function in the sense that it kills both the scalar 
curvature and the traceless Ricci tensor of $g_1$ moreover blows up 
sufficiently fast along the exotic $\R^4$-end of $X_M$ to make the 
rescaled metric $g_0$ complete. A classical example serves as a 
motivation. Put $(\hat{X}_M,\hat{g}_1):=(S^4,\hat{g}_1)$ that is, the 
$4$-sphere $S^4\subset\R^5$ equipped with its standard orientation and 
round metric $\hat{g}_1$ inherited from the embedding. It is well-known 
that $(S^4,\hat{g}_1)$ is self-dual and Einstein with non-zero cosmological 
constant i.e., not Ricci-flat. Put $X_M:=S^4\setminus\{\infty\}\cong\R^4$ to 
be the standard $\R^4$; then $g_1:=\hat{g}_1\vert_{\R^4}$ is an incomplete 
self-dual metric on $\R^4$ but picking $\varphi (r):=\frac{1}{1+r^2}$ where 
$r$ is the radial coordinate on $\R^4$ from its origin i.e., $\varphi$ 
vanishes exactly in $\{\infty\}\in S^4$, then $g_0:=\varphi^{-2}\cdot g_1$ 
is nothing but the standard flat metric on $\R^4$ which is of course complete 
and Ricci-flat. Hence $(X_M,g_0):=(\R^4, \varphi^{-2}g_1)$, the 
conformal rescaling of $(X_M,g_1)=(\R^4,g_1)$, is the desired complete 
Ricci-flat space in this simple case. In our much more general situation 
we shall use Penrose' non-linear graviton construction (i.e., twistor theory) 
\cite{pen2} to find conformal rescalings. 

\begin{remark}\rm Let us first 
recall Penrose' twistor method to solve the Riemannian vacuum Einstein 
equation (for a very clear introduction cf. \cite{hit1,hit2}). 
Consider the projectivized negative chiral spinor bundle $P(\hat{\Sigma}^-)$ 
over for instance the compact self-dual space $(\hat{X}_M,\hat{g}_1)$ in 
Lemma \ref{kezdolemma}; note that this bundle exists even if $\hat{X}_M$ is 
not spin. Since in $4$ dimensions $\hat{\Sigma}^-$, if exists, is a rank 
$2$ complex vector bundle over $\hat{X}_M$, its projectivization 
$P(\hat{\Sigma}^-)$ is the total space of a smooth $\C P^1$-fibration 
$\hat{p}:P(\hat{\Sigma}^-)\rightarrow \hat{X}_M$. The 
Levi--Civita connection of any metric on $\hat{X}_M$ can be used to 
furnish the real $6$-manifold $P(\hat{\Sigma}^-)$ with a canonical almost 
complex structure; the fundamental observation of twistor theory is that this 
almost complex structure now is integrable because $\hat{g}_1$ is 
self-dual. The resulting complex $3$-manifold $\hat{Z}\cong 
P(\hat{\Sigma}^-)$ is called the {\it twistor space} while the smooth fibration 
$\hat{p}:\hat{Z}\rightarrow \hat{X}_M$ the {\it twistor fibration} of 
$(\hat{X}_M,\hat{g}_1)$. The most 
important property of a twistor space of this kind is that its twistor 
fibers $\hat{p}^{-1}(x)\subset \hat{Z}$ for all $x\in \hat{X}_M$ fit into 
a locally complete complex $4$-paremeter family $\hat{X}_M^\C$ of 
projective lines $Y\subset\hat{Z}$ each with normal bundle $H\oplus H$, 
with $H$ being the dual of 
the tautological line bundle over $Y\cong\C P^1$. Moreover, there exists a 
real structure $\hat{\tau} :\hat{Z}\rightarrow\hat{Z}$ defined by taking the 
antipodal maps along the twistor fibers $\C P^1\cong\hat{p}^{-1}(x)\subset 
\hat{Z}$ for all 
$x\in\hat{X}_M\subset \hat{X}_M^\C$ which are therefore called ``real 
lines'' among all the lines in $\hat{X}_M^\C$. In other words, $\hat{Z}$ 
is fibered exactly by the real lines $Y_x:=\hat{p}^{-1}(x)$ for all $x\in 
\hat{X}_M$. Hence the real $4$ 
dimensional self-dual geometry has been encoded into a $3$ dimensional complex 
analytic structure in the sense that one can recover 
$(\hat{X}_M,\hat{g}_1)$ just from $\hat{Z}$ up to conformal equivalence. 

One can go further and raise the question how to recover precisely 
$(\hat{X}_M, \hat{g}_1)$ itself from its conformal class, or more 
interestingly to us: how to get a Ricci-flat Riemannian $4$-manifold 
$(X_M,g_0)$ i.e., a solution of the (self-dual) Riemannian vacuum Einstein 
equation. Not surprisingly, to get the 
latter stronger structure, one has to specify further data on the twistor 
space. A fundamental result of twistor theory is that a solution of the 
$4$ dimensional (self-dual) Riemannian vacuum Einstein equation is 
equivalent to the following set of data (cf. \cite{hit1,hit2}):

\begin{itemize}

\item[$*$] A complex $3$-manifold $Z$, the total space of
a holomorphic fibration $\pi :Z\rightarrow\C P^1$;

\item[$*$] A complex $4$-paremeter family of holomorphically 
embedded complex projective lines $Y\subset Z$, each with normal bundle 
$H\oplus H$ (here $H$ is the dual of the tautological
bundle i.e., the unique holomorphic line bundle on $Y\cong\C P^1$ with 
$\langle c_1(H), [Y]\rangle =1$);

\item[$*$] A non-vanishing holomorphic section $s$ of
$K_Z\otimes\pi^*H^4$ (here $K_Z$ is the canonical bundle of $Z$);

\item[$*$] A real structure $\tau :Z\rightarrow Z$ such that 
$Z$ is fibered by the $\tau$-invariant elements $Y\subset Z$ of the 
family (these are called ``real lines'') and $\tau$ coincides 
with the antipodal map $u\mapsto -\overline{u}^{-1}$ upon restricting 
to the real lines; moreover $\pi$ and $s$ are compatible with $\tau$. 

\end{itemize}

\noindent These data allow one to construct a Ricci-flat 
and self-dual (i.e., the Ricci tensor and the anti-self-dual part of 
the Weyl tensor vanishes) solution $(X_M,g_0)$ of the {\it Riemannian} 
Einstein's vacuum equation with vanishing cosmological constant as 
follows. The holomorphic lines $Y\subset Z$ form a locally complete 
family and fit together into a complex $4$-manifold $X_M^\C$. This space 
carries a natural complex conformal structure by declaring two nearby points 
$y_1,y_2\in X_M^\C$ to be null-separated if the corresponding lines 
intersect i.e., $Y_1\cap Y_2\not=\emptyset$ in $Z$. Infinitesimally this 
means that on every tangent space $T_yX_M^\C =\C^4$ a null cone is 
specified. Restricting this to the real lines singled out by 
$\tau$ and parameterized by an embedded real 
$4$-manifold $X_M\subset X_M^\C$ we obtain the real conformal class $[g_0]$ 
of a Riemannian metric on $X_M$. The isomorphism $s: K_Z\cong\pi^*H^{-4}$ 
is essentially uniquely fixed by its compatibility with $\tau$ and $\pi$ and 
gives rise to a volume form on $X_M$ this way fixing the metric $g_0$ in 
the conformal class. Given the conformal class, it is already meaningful 
to talk about the projectivized negative chiral spinor bundle 
$P(\Sigma^-)$ over $X_M$ with its induced orientation from the twistor 
space and $Z$ can be identified with the total space of $P(\Sigma^-)$. This 
way we obtain a smooth twistor fibration $p:Z\rightarrow X_M$ whose fibers 
are $\C P^1$'s hence $\pi :Z\rightarrow\C P^1$ can be regarded as a parallel 
translation along this bundle over $X_M$ with respect to a flat connection 
which is nothing but the induced negative spin connection of $g_1$ on 
$\Sigma^-$. Knowing the decomposition of the Riemannian curvature into 
irreducible components over an oriented Riemannian $4$-manifold 
\cite{sin-tho}, this partial 
flatness of $P(\Sigma^-)$ implies that $g_0$ is Ricci-flat and self-dual. 
Finally note that, compared to the bare twistor space $\hat{Z}$ of a 
self-dual manifold $(\hat{X}_M,\hat{g}_1)$ above, the essential new 
requirement for constructing a self-dual {\it Ricci-flat} space $(X_M,g_0)$ 
is the existence of a holomorphic map $\pi$ from the twistor space $Z$ into 
$\C P^1$. We conclude our summary of the non-linear graviton construction by 
referring to \cite{hit1, hit2} for further details.
\end{remark}

\noindent In the case of our situation set up in Lemma \ref{kezdolemma} 
twistor theory works as follows. Consider the compact self-dual 
space $(\hat{X}_M,\hat{g}_1)$ from Lemma \ref{kezdolemma}, take its 
twistor fibration $\hat{p}: \hat{Z}\rightarrow \hat{X}_M$ and let 
\[p:\:\:Z\longrightarrow X_M\] 
be its restriction induced by the smooth embedding $X_M\subsetneqq 
\hat{X}_M$ i.e., $Z:=\hat{Z}\vert_{X_M}$ and 
$p:=\hat{p}\vert_{X_M}$. Then $Z$ is a non-compact 
complex $3$-manifold already obviously possessing all the required twistor 
data except the existence of a holomorphic mapping $\pi :Z\rightarrow\C P^1$. 

\begin{lemma} Consider $(X_M,g_1)$ as in Lemma \ref{kezdolemma} 
with its twistor fibration $p:Z\rightarrow X_M$ constructed above. If 
$\pi_1(X_M)=1$ (i.e., the original compact manifold satisfies $\pi_1(M)=1$) 
then there exists a holomorphic mapping $\pi :Z\rightarrow\C P^1$.

\label{pilemma}
\end{lemma}

\noindent {\it Proof.} Let $x_0\in X_M$ be a fixed point belonging to 
the exotic $\R^4$-factor $R^4$ of $X_M$ in its decomposition 
(\ref{ujsokasag}). Our aim is to construct a holomorphic map 
\begin{equation}
\pi:\:\:Z\longrightarrow p^{-1}(x_0)\cong\C P^1
\label{celfuggveny}
\end{equation}
that we carry out in three steps.

Firstly over an exotic $R^4\subset \C P^2$ we construct by classical means 
holomorphic maps parameterized by ideal points $x\in\C P^2\setminus R^4$. It 
is known that $\hat{Z}(\C P^2)\cong P(T^*\C P^2)$ i.e., the twistor space of 
the complex projective space can be identified 
with its projective cotangent bundle. Consequently $\hat{Z}(\C P^2)$ can 
be described as the flag manifold $F_{12}(\C ^3)$ consisting of pairs 
$({\mathfrak l},{\mathfrak p})$ where $0\in{\mathfrak l}\subset\C^3$ is a 
line and ${\mathfrak l}\subset{\mathfrak p}\subset\C^3$ is a plane containing 
the line. Then in the twistor fibration $\hat{p}:\hat{Z}(\C P^2)
\rightarrow\C P^2$ of the complex projective space $\hat{p}$ sends 
$({\mathfrak l},{\mathfrak p})\in F_{12}(\C ^3)$ into the point 
$[\:{\mathfrak l}\:]\in\C P^2$ corresponding to ${\mathfrak l}\subset\C^3$. 
This is a smooth $\C P^1$-fibration over $\C P^2$. Part (i) of Theorem 
\ref{egzotikusnagycsalad} tells us that $R^4\subset\C P^2$. Writing 
$Z(R^4):=\hat{Z}(\C P^2)\vert_{R^4}$ and $p:=\hat{p}\vert_{R^4}$ the 
restricted twistor fibration $p:Z(R^4)\rightarrow R^4$ is topologically 
trivial i.e., $Z(R^4)$ is homeomorphic to $R^4\times S^2\cong\R^4\times 
S^2$ because $R^4$ is contractible.\footnote{This is a necessary 
topological condition for the existence of the map (\ref{celfuggveny}). 
The full twistor fibration $\hat{p}:\hat{Z}(\C P^2)\rightarrow\C P^2$ is 
non-trivial, neither its restriction to the punctured space 
$\C P^2\setminus\{x\}$.} Take a starting pair 
$({\mathfrak l},{\mathfrak p})\in Z(R^4)$ with a running point 
$[{\mathfrak l}]\in R^4$. Fix a target point $[{\mathfrak l}_0]\in R^4$ with 
$p^{-1}([{\mathfrak l}_0])\subset Z(R^4)$ consisting of terminating pairs 
$({\mathfrak l}_0,{\mathfrak p}_0)$. Fix an ideal point 
$x\in\C P^2\setminus R^4$ hence surely $x\not=[{\mathfrak l}]$ as well as 
$x\not=[{\mathfrak l}_0]$.  

Now we construct a map $\pi_x:Z(R^4)\rightarrow 
p^{-1}([{\mathfrak l}_0])\cong\C P^1$ as follows. By the aid of the 
Fubini--Study metric one can to talk about distances and angles on $\C P^2$. 
Then surely $d([{\mathfrak l}],x)>0$ therefore there exists a unique 
projective line in $\C P^2$ passing through 
$[{\mathfrak l}]$ and $x$ and precisely {\it two} perpendicular 
bisectors of the corresponding two segments along this projective 
line.\footnote{Recall how to construct them. First let $y\in\C P^2$ 
be a midpoint along one segment connecting $[{\mathfrak l}]$ and 
$x$ i.e., satisfying $d([{\mathfrak l}],y)=d(y,x)=
\frac{1}{2}d([{\mathfrak l}],x)$; second viewing the distinct points 
$[{\mathfrak l}]$ and $x$ in $\C P^2$ as two linearly independent vectors in 
$\C^3$ let $y_\infty\in\C P^2$ be the point corresponding to their complex 
vectorial product in $\C^3$. Then $y_\infty$ is not on the segment hence 
$y_\infty\not=y$ and a perpendicular 
bisector arises by taking the unique projective line in $\C P^2$ 
through $y$ and $y_\infty$.} 
Let $\ell\subset\C P^2$ be one continuous choice of these perpendicular 
bisectors as $[{\mathfrak l}]\in R^4$ varies. Now, take 
$({\mathfrak l},{\mathfrak p})\in p^{-1}([{\mathfrak l}])\subset Z(R^4)$; 
there is a unique intersection point $[{\mathfrak p}]\cap\ell\in\C P^2$ 
and consider the unique line $m\subset\C P^2$ connecting 
$[{\mathfrak p}]\cap\ell$ and the ideal point $x$. We denote this 
operation so far as $P_x([{\mathfrak p}])=m$. Next, since 
$d(x,[{\mathfrak l}_0])>0$, we can repeat the whole procedure replacing 
the running $[{\mathfrak l}]\in R^4$ with the fixed target point 
$[{\mathfrak l}_0]\in R^4$. That is, let $\ell_0\subset\C P^2$ be a fixed 
perpendicular bisector of the line through $x$ and $[{\mathfrak l}_0]$; 
then there is a unique point $m\cap\ell_0$ and finally, define the pair 
$({\mathfrak l}_0,{\mathfrak p}_0)\in p^{-1}([{\mathfrak l}_0])$ such that 
$[{\mathfrak p}_0]\subset\C P^2$ is the unique line connecting 
$m\cap\ell_0$ with $[{\mathfrak l}_0]\in R^4$. Again, denote this 
operation by $R_x(m)=[{\mathfrak p}_0]$. In short, 
\begin{equation}
\pi_x(({\mathfrak l},{\mathfrak p})):=
({\mathfrak l}_0, {\mathfrak p}_0)
\mbox{where ${\mathfrak p}_0\subset\C^3$ is the line 
$[{\mathfrak p}_0]\subset\C P^2$ satisfying $R_x(P_x([{\mathfrak p}]))=
[{\mathfrak p}_0]$}
\label{pix}
\end{equation}
(see Figure 2 for a construction of this map in 
projective geometry). It is a classical observation that this map is 
well-defined and holomorphic; in particular it is the identity on 
$p^{-1}([{\mathfrak l}_0])\subset Z(R^4)$ i.e., 
$\pi_x(({\mathfrak l}_0,{\mathfrak p}_0))=
({\mathfrak l}_0, {\mathfrak p}_0)$. For clarity we note that 
$\pi_x:Z(R^4)\rightarrow p^{-1}([{\mathfrak l}_0])$ is single-valued along 
$R^4$ in spite of the fact that $\pi_x$ itself extends over the 
larger punctured space $\C P^2\setminus\{x\})\supset R^4$ and is 
double-valued there due to the ambiguity in the choice of the perpendicular 
bisector; however fortunately one cannot pass continuously to another branch of 
$\pi_x$ without crossing somewhere the infinitely distant $2$-sphere 
$\C P^2\setminus R^4\subset\C P^2$. 
\vspace{0.7in}

\centerline{
\setlength{\unitlength}{1cm}
\begin{picture}(5,3)
\thicklines
\put(1,1){\circle*{0.2}}
\put(1.2,0.8){$[\:{\mathfrak l}\:]$}
\put(3,3){\circle*{0.2}}
\put(3,3.2){$x$}
\put(5,1){\circle*{0.2}}
\put(4.1,0.8){$[\:{\mathfrak l}_0\:]$}
\put(3.5,0.5){\line(-1,1){3}}
\put(0.2,3.5){$\ell$}
\put(2.5,0.5){\line(1,1){3}}
\put(5.7,3.5){$\ell_0$}
\put(1,0.5){\line(0,1){3}}
\put(0.5,1.5){$[{\mathfrak p}]$}
\put(5,0.5){\line(0,1){3}}
\put(5.2,1.5){$[{\mathfrak p}_0]$}
\put(0.5,3){\line(1,0){5}}
\put(2,2.6){$m$}
\put(7,2){$\C P^2$}
\end{picture}
            }
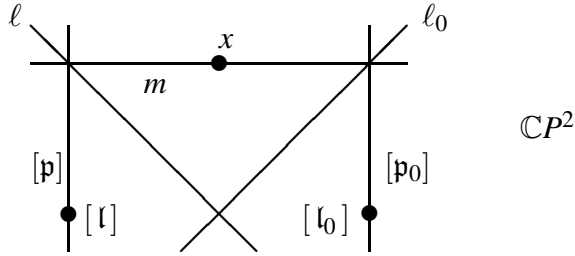
\centerline{Figure 2. Two-step construction of the map $\pi_x$ 
satisfying $\pi_x(({\mathfrak l},{\mathfrak p}))=
({\mathfrak l}_0,{\mathfrak p}_0)$.}
\vspace{0.3in}

Secondly we fuse all the maps $\pi_x:Z(R^4)\rightarrow 
p^{-1}([{\mathfrak l}_0])$ in (\ref{pix}), when $x\in\C P^2\setminus R^4$ 
runs through the complement of the exotic $\R^4$, into a single-valued 
holomorphic map $\pi:Z(R^4)\rightarrow p^{-1}([{\mathfrak l}_0])$ by applying 
the concept of Lebesgue integration of algebraic-function-field-valued 
functions summarized in the Appendix. Assume that with a fixed ideal point 
$x\in\C P^2\setminus R^4$ the holomorphic map (\ref{pix}) is given; take 
now a {\it different} ideal point $y\in\C P^2\setminus R^4$ with its 
corresponding holomorphic map 
$\pi_y:Z(R^4)\rightarrow p^{-1}([{\mathfrak l}_0])$ into the {\it same} 
target space. Then there exists a commutative diagram 
\[\xymatrix{
 \ar[dr]_{\pi_y}Z(R^4)\ar[r]^{\pi_x} & 
p^{-1}([{\mathfrak l}_0])\ar[d]^{f_{yx}} \\
                   & p^{-1}([{\mathfrak l}_0])}\] 
with $f_{yx}$ being a holomorphic map satisfying $f_{xx}={\rm 
Id}_{p^{-1}([{\mathfrak l}_0])}$. Pick 
an affine coordinate system $(u,v)$ on a coordinate ball $U\subset\C P^2$ 
centered about $[{\mathfrak l}_0]\in U$ i.e., $(u([{\mathfrak l}_0]),
v([{\mathfrak l}_0])=(0,0)$. In this coordinate system any 
affine line $[{\mathfrak p}_0]\cap U$ passing through $[{\mathfrak l}_0]$ 
looks like $(u([{\mathfrak p}_0]), v([{\mathfrak p}_0]))=(u,zu)$ with 
$z\in\C\cup\{\infty\}=\C P^1$ hence $({\mathfrak l}_0,{\mathfrak p}_0)=z$ 
provides us with an identification 
$p^{-1}([{\mathfrak l}_0])\cong\C P^1$. However it is known for a 
long time that a holomorphic map from $\C P^1$ into itself is a 
rational function in a single variable; consequently under this 
identification $f_{yx}: p^{-1}([{\mathfrak l}_0])\rightarrow 
p^{-1}([{\mathfrak l}_0])$ can be described by a unique element $R_{yx}$ 
of the algebraic function field $\C(z)$, the complex rational functions in 
one variable $z$, satisfying $R_{xx}(z)=z$. That is, there 
exist complex-coefficient polynomials 
$P_{yx}(z)=a_m(y)z^m +\dots +a_1(y)z+a_0(y)$ and 
$Q_{yx}(z)=b_n(y)z^n+\dots +b_1(y)z+b_0(y)$ such that 
$R_{yx}(z)=\frac{P_{yx}(z)}{Q_{yx}(z)}$ and $R_{xx}(z)=z$ implies 
$a_1(x)=1$ and $b_0(x)=1$ and all the rest being zero at $x$. In this 
context for a fixed $({\mathfrak l},{\mathfrak p})\in Z(R^4)$ it is worth 
regarding $\pi_x(({\mathfrak l},{\mathfrak p}))$ as a particular choice for 
$z$ in the abstractly given algebraic function field $\C (z)$ 
and denoting this coordinatized $(\C (z),\pi_x)$ simply as 
$\C (\pi_x)$. We eventually come up with 
\[\pi_y (({\mathfrak l},{\mathfrak p})) =
R_{yx}(\pi_x(({\mathfrak l},{\mathfrak p})))= 
\frac{P_{yx}(\pi_x(({\mathfrak l},{\mathfrak p})))}
{Q_{yx}(\pi_x(({\mathfrak l},{\mathfrak p})))}\] 
and the coefficients $a_i, b_j:\C P^2\setminus R^4\rightarrow\C$ of 
$P_{yx}$ and $Q_{yx}$ respectively, are at least continuous functions 
assuming perhaps zero values, therefore the degrees of $P_{yx}$ and 
$Q_{yx}$ can jump as $y$ runs 
through the ideal points. Nevertheless, exploiting the compactness of 
$\C P^2\setminus R^4$ (homeomorphic to $S^2$) and the continuity of the 
coefficients, one can see that there exist overall constants $N_x\in\N$ 
and $K_x\in\R^+$ such that 
\[\max\left(\sup\limits_{y\in\C P^2\setminus R^4}\deg P_{yx}\:,\: 
\sup\limits_{y\in\C P^2\setminus R^4}\deg Q_{yx}\right)\leqq 
N_x\:\:\:,\:\:\: \max\limits_{0\leqq i,j\leqq N_x}\left(\sup\limits_{y\in\C 
P^2\setminus R^4} \vert a_i(y)\vert\:,\:\sup\limits_{y\in\C P^2\setminus 
R^4}\vert b_j(y)\vert\right)\leqq K_x\:\:\:.\] 
Let $S^2\subset\R^3$ denote the standard $2$-sphere with its inherited 
orientation, smooth structure and round metric and let 
$i: S^2\rightarrow\C P^2$ be a continuous embedding such that 
$i:S^2\rightarrow\C P^2\setminus R^4$ is a homeomorphism onto 
the complement. In this way the coefficients of $P_{yx}$ and $Q_{yx}$ give 
rise to continuous functions on the standard $2$-sphere via pullback and 
we obtain a continuous function $i^*y\mapsto R_{i^*y,x}$  from $S^2$ into 
$\C (\pi_x)$. Writing $\dd (i^*y)$ for the usual volume-form on $S^2$ with 
respect to its orientation and round metric we define $\pi:Z(R^4)\rightarrow 
p^{-1}([{\mathfrak l}_0])$ by 
\begin{equation}
\pi(({\mathfrak l},{\mathfrak p})):=\int\limits_{S^2}
R_{i^*y,x}(\pi_x(({\mathfrak l},{\mathfrak p})))\:\dd (i^*y)
\label{azintegral}
\end{equation}
for all $({\mathfrak l},{\mathfrak p})\in Z(R^4)$. As explained in the 
Appendix, the expression on the right hand side as an 
algebraic-function-field-valued Lebesgue integral over $S^2$ exists moreover 
the map $({\mathfrak l},{\mathfrak p})\mapsto
\pi(({\mathfrak l},{\mathfrak p}))$ in (\ref{azintegral}) is holomorphic and 
is independent of $x$; these are proved in Lemma 
\ref{holomorflemma}. In particular we can think 
of $\int_{S^2}R_{i^*y,x}(\pi_x(\divideontimes))\:\dd (i^*y)\in\C (\pi_x)$ as 
a rational function in the variable $\pi_x$ and changing the reference 
point $x$ just corresponds to using different coordinatizations in the 
abstract function field $\C (z)$. 

Thirdly we extend the map (\ref{azintegral}) over the whole $X_M$. 
Let $y_0\in K\subset R^4$ be the attaching point used to glue $R^4$ with the 
rest of $X_M$ as in Lemma \ref{kezdolemma}; we suppose 
$y_0\not=[{\mathfrak l}_0]\in R^4$. Let 
$j:R^4\setminus\{y_0\}\rightarrow X_M$ be a smooth embedding 
which identifies $R^4$ with the exotic $\R^4$-end of $X_M$ in its 
decomposition (\ref{ujsokasag}) such that $j([{\mathfrak l}_0])=x_0$ 
where $x_0\in X_M$ is the distinguished point of the map (\ref{celfuggveny}) 
to be constructed. Also write $J:Z(R^4\setminus\{y_0\})\rightarrow Z$ 
for the induced inclusion of the twistor space into that of $X_M$. Then 
$\pi':=(J^{-1})^*(\pi\vert_{Z(R^4\setminus\{y_0\})})\::\: V\rightarrow 
p^{-1}(x_0)$ is a partially defined holomorphic map on a connected open subset 
$V:=p^{-1}(j(R^4\setminus\{y_0\}))\subset Z$ of the twistor space of $X_M$. 
We now extend $\pi'$ holomorphically over the whole $Z$ to be the map 
(\ref{celfuggveny}) as follows. Consider an open covering $X_M=\cup_kU_k$ 
giving rise to an open covering $Z=\cup_kp^{-1}(U_k)$ of the twistor 
space, too. If $x\in X_M$ is an inner point of the exotic $\R^4$-end 
$j(R^4\setminus\{y_0\})\subset X_M$ 
(but different from the base point $x_0$) such that an open subset 
$x\in U_k$ from the covering satisfies $U_k\subset 
j(R^4\setminus\{y_0\})\subset X_M$ then we get a neighbourhood 
$p^{-1}(U_k)\subset V\subset Z$ of $p^{-1}(x)$, too. We know that the 
holomorphic map $\pi'\vert_{p^{-1}(x)}:p^{-1}(x)\rightarrow p^{-1}(x_0)$ 
extends to a holomorphic map $\pi'\vert_{p^{-1}(U_k)}:p^{-1}(U_k)\rightarrow 
p^{-1}(x_0)$. 
However, by referring at this step to an important extendibility result of 
Griffiths \cite[Proposition 1.3]{gri}, this extendibility depends only on two 
holomorphic data: the pullback tangent bundle 
$(\pi'\vert_{p^{-1}(x)})^*(Tp^{-1}(x_0))$ over $p^{-1}(x)$ and the normal 
bundle of it as a complex submanifold $p^{-1}(x)\subset Z$. 
But the former bundle cannot locally depend on $x$ because holomorphic line 
bundles over $p^{-1}(x)\cong\C P^1$ form a discrete set. Regarding the 
latter bundle, $p^{-1}(x)\subset Z$ as a submanifold is a {\it twistor line} 
in $Z$ and all twistor lines in the twistor fibration have isomorphic 
normal bundles (see the {\it Remark} on twistor theory above). Since these 
twistor lines fulfill the whole $Z$ these arguments convince us that using 
the open covering $\cup_kp^{-1}(U_k)$ of $Z$ and exploiting the simply 
connectedness of $Z$ provided by that of $X_M$ we can analytically continue 
the partial map $\pi'$ above from the connected open subset $V\subset Z$ to a 
holomorphic map (\ref{celfuggveny}) over the whole $Z$  in a unique way as 
desired. $\Diamond$  
\vspace{0.1in}

\noindent It also follows that $\pi :Z\rightarrow\C P^1$ i.e., the map 
(\ref{celfuggveny}) constructed in Lemma \ref{pilemma} is compatible with the 
real structure $\tau :Z\rightarrow Z$ already fixed by the self-dual structure 
in Theorem \ref{taubestetel} therefore twistor theory provides us with a 
Ricci-flat (and self-dual) Riemannian metric $g_0$ on $X_M$. We proceed 
further and demonstrate that, unlike $(X_M,g_1)$, the space $(X_M,g_0)$ is 
complete. 
\begin{lemma}
The Ricci-flat Riemannian manifold $(X_M,g_0)$ is complete. 
\label{teljeslemma}
\end{lemma}

\begin{remark}\rm Moreover $(X_M,g_0)$ is simply connected and self-dual i.e., 
as a by-product, is in fact a hyper-K\"ahler space. In particular 
taking $M:=S^4$ then $\hat{X}_{S^4}=\C P^2$ so $X_{S^4}=R^4$, the 
largest member of the Gompf--Taubes radial family, carries a complete 
hyper--K\"ahler metric. Hence these spaces are {\it gravitational 
instantons} with dominant contribution to the Euclidean quantum gravity 
path integral \cite{ass,dus}.
\end{remark} 

\noindent {\it Proof of Lemma \ref{teljeslemma}}. Since both $g_1$ and 
this Ricci-flat metric $g_0$ stem from the same complex structure on the same 
twistor space $Z$ we know from twistor theory that these metrics are in fact 
conformally equivalent. That is, there exists a smooth non-constant strictly 
positive function $\varphi :X_M\rightarrow\R^+$ such that 
$\varphi^{-2}\cdot g_1= g_0$. Our strategy to prove completeness is to 
follow Gordon \cite{gor} i.e., to demonstrate that an appropriate 
real-valued function on $X_M$, in our case $\log\varphi^{-1}:X_M\rightarrow\R$, 
is proper (i.e., the preimages of compact subsets are compact) with bounded 
gradient in modulus with respect to $g_0$ implying the completeness.        

Referring to (\ref{ujsokasag}) the open space $X_M$ arises by deleting 
$\C P^2\setminus R^4$ from a $\C P^2$-factor of the closed 
space $\hat{X}_M$.
First we observe that $\varphi^{-1}:X_M\rightarrow\R^+$ is uniformly 
divergent along $\C P^2\setminus R^4$ as follows. It is clear that the 
potential singularities of $\varphi^{-1}$ stem from those of the map 
(\ref{celfuggveny}). The map $\pi_x$ constructed in (\ref{pix}) has an 
obvious singularity at $x\in\C P^2\setminus R^4$ and $\pi$ itself has been 
constructed in (\ref{azintegral}) by integrating together all the $\pi_x$'s 
along $\C P^2\setminus R^4$ consequently $\pi$ is singular along the whole 
$\C P^2\setminus R^4$. Consequently $\varphi^{-1}$ is expected to 
be somehow singular along the whole $\C P^2\setminus R^4$, too. Moreover, this 
part of the construction of $\pi$ in Lemma \ref{pilemma} deals with a single 
$\C P^2$-factor in (\ref{ujsokasag}) only hence is universal in the sense that 
it is independent of the $M$-factor in (\ref{ujsokasag}). In other words, 
for all $X_M$ the map (\ref{celfuggveny}) arises by analytically 
continuing the {\it same} $\pi$ on $R^4$ constructed in the first two steps 
in Lemma \ref{pilemma}. So we anticipate $\varphi^{-1}:X_M\rightarrow\R^+$ with 
$X_M\subset\hat{X}_M$ to possess a uniform and universal singular 
behaviour along $\C P^2\setminus R^4\subset\hat{X}_M$ what we analyze now 
further. 

The conformal scaling function satisfies with respect to 
$g_1$ the following equations on $X_M$:
\begin{equation}
\left\{\begin{array}{ll}
\Delta\varphi^{-1}+\frac{1}{6}\varphi^{-1}{\rm Scal}_1&=0\:\:\:
\mbox{(vanishing of the scalar curvature of $g_0$ on $X_M$)};\\
&\\
\nabla^2\varphi -\frac{1}{4}\Delta\varphi\cdot
g_1+\frac{1}{2}\varphi\cdot {\rm Ric}_1^0&= 0\:\:\:
\mbox{(vanishing of the traceless Ricci tensor of $g_0$ on $X_M$)}.
\end{array}\right.
\label{skalazas}
\end{equation}
The Ricci tensor ${\rm Ric}_1$ of $g_1$ extends smoothly over $\hat{X}_M$ 
because it is just the restriction of the Ricci tensor of the self-dual 
metric $\hat{g}_1$ on $\hat{X}_M$. Therefore both its scalar curvature 
${\rm Scal}_1$ and traceless Ricci part ${\rm Ric}^0_1$ extend. 
Consequently from the first equation of (\ref{skalazas}) we can see that 
$\varphi\Delta\varphi^{-1}$ extends smoothly over $\hat{X}_M$. Likewise, adding 
the tracial part to the second equation of (\ref{skalazas}) we get 
$\varphi^{-1}\nabla^2\varphi =-\frac{1}{2}{\rm Ric}_1$ hence we conclude 
that the symmetric tensor field $\varphi^{-1}\nabla^2\varphi$ extends 
smoothly over $\hat{X}_M$ so its trace $\varphi^{-1}\Delta\varphi$ as well. 
The equation $\Delta (\varphi\cdot\varphi^{-1})=0$ gives the standard 
identity $0=(\Delta\varphi) \varphi^{-1} +2g_1(\dd\varphi\:,\dd\varphi^{-1})+
\varphi\Delta\varphi^{-1}$ and adjusting this a bit we get
\begin{equation}
\varphi^2\vert\dd\varphi^{-1}\vert^2_{g_1}=
\frac{1}{2}(\varphi\Delta\varphi^{-1}+\varphi^{-1}\Delta\varphi )
\label{azonossag}
\end{equation} 
consequently the function $\varphi\vert\dd\varphi^{-1}\vert_{g_1}$ 
extends smoothly over $\hat{X}_M$, too. Assume now that $\varphi^{-1}$ is 
extendible over $\hat{X}_M$ at least continuously. Then, 
taking into the aforementioned universal behaviour of $\varphi^{-1}$ around 
its interesting part, we can take $\hat{X}_{S^4}=\C P^2$ and 
$\hat{g}_1=\mbox{Fubini--Study metric}$. However this metric has constant 
scalar curvature consequently, by the aid of the first equation of 
(\ref{skalazas}) and the maximum principle, we could conclude that 
$\varphi^{-1}$ is constant on $\C P^2$, a contradiction. Assume now that 
$\varphi^{-1}$ does not extend continuously over $\hat{X}_M$ but 
$\vert\varphi^{-1}\vert$ is bounded. Then its 
gradient $\dd\varphi^{-1}$ gets diverge along $\C P^2\setminus R^4$ 
hence from the extendibility of $\varphi\vert\dd\varphi^{-1}\vert_{g_1}$ we 
obtain that $\varphi$ vanishes along $\C P^2\setminus R^4$, a contradiction 
again. Therefore $\varphi^{-1}:X_M\rightarrow\R^+$ with $X_M\subset 
\hat{X}_M$ is uniformly divergent along $\C P^2\setminus R^4\subset 
\hat{X}_M$ yielding, on the one hand, that $\log\varphi^{-1}:X_M\rightarrow\R$ 
is a proper function. 
 
As a by-product the inverse function $\varphi$ is bounded on $X_M$ i.e., 
$\vert\varphi\vert\leqq c_1$ with a finite constant. 
We already know that $\vert\varphi\Delta\varphi^{-1}\vert\leqq c_2$ and 
$\vert\varphi^{-1}\Delta\varphi\vert\leqq c_3$ with other finite 
constants as well. 
Since $\varphi\vert\dd\varphi^{-1}\vert_{g_1}=\vert\dd 
(\log\varphi^{-1})\vert_{g_1}$ and carefully noticing that 
$\vert\xi\vert_{g_0}=\varphi\vert\xi\vert_{g_1}$ on
$1$-forms we can use (\ref{azonossag}) and the estimates above to come up with 
\[\vert\dd (\log \varphi^{-1})\vert^2_{g_0}
\leqq c_1^2\vert\dd (\log \varphi^{-1})\vert^2_{g_1}
\leqq c_1^2\left(\left\vert\varphi\Delta\varphi^{-1}\vert 
+\vert\varphi^{-1}\Delta\varphi\right\vert\right)\leqq c_1^2(c_2+c_3)<+\infty\]
and conclude, on the other hand, that $\log\varphi^{-1}: 
X_M\rightarrow\R$ has bounded gradient in modulus with respect to $g_0$. 
Therefore, in light of Gordon's theorem \cite{gor}, the Ricci-flat space 
$(X_M,g_0)$ is complete. $\Diamond$
\vspace{0.1in}

\noindent {\it Proof of Theorem
\ref{riemannverzio}}. The proof now readily follows by putting together 
Lemmata \ref{kezdolemma}, \ref{pilemma} and \ref{teljeslemma}. $\Diamond$ 
\vspace{0.1in}


\section{Lorentzian considerations}
\label{four}


Having established the existence of an abundance of spaces, 
it is worth summarizing the situation before 
we proceed further. In Section \ref{three} we have constructed 
certain non-compact complete Ricci-flat Riemannian $4$-manifolds. 
These geometries are hyper-K\"ahler as a by-product however, more 
important to us, they have the odd feature that---although they are 
non-compact---surely do not split smoothly into the product of any 
$3$-manifold and the real line (with their unique smooth structures) 
because they contain a ``creased'' asymptotical region, more precisely a 
single end diffeomorphic to an exotic $\R^4$ (see Figure 1). Taking 
into account \cite{ber-san, che-nem} this non-splitting phenomenon offers a 
good starting point to violate the {\bf SCCC} in a generic way. However, 
our solutions of the vacuum Einstein equation are still {\it Riemannian} 
hence we have to work on them further to obtain solutions of the {\it 
Lorentzian} vacuum Einstein equation on the same ``creased'' manifolds. In 
this section we will prove the following theorem whose proof again needs some 
preparations and will be presented at the end of this section.
\begin{theorem}

Consider the Riemannian $4$-manifold $(X_M,g_0)$ as in Theorem 
\ref{riemannverzio}. Then out of this space one can construct an oriented 
smooth Lorentzian $4$-manifold 
\[(X_M,g)\]
with the following properties. 

The metric $g$ is a Ricci-flat, probably null and-or timelike geodesically 
incomplete, but surely not globally hyperbolic metric on $X_M$. Furthermore 
if $(S,h)\subset (X_M,g)$ is any connected, oriented, complete spacelike 
sub-$3$-manifold with corresponding (necessarily partial) initial data set 
$(S,h,k)$, then any sufficiently large perturbation $(X_M',g')$ of $(X_M,g)$ 
relative to $(S,h,k)$ in the sense of Definition \ref{perturbacio} is not 
globally hyperbolic. Here ``sufficiently large'' means that $X'_M$, 
satisfying $S\subset X'_M\subseteqq X_M$, contains the image, present in 
the $R^4$-factor of $X_M$ in its decomposition (\ref{ujsokasag}), of the 
compact subset $K\subset R^4$ of part (i) in Theorem \ref{egzotikusnagycsalad}.
\label{lorentzverzio}
\end{theorem}

\noindent Take a Riemannian $4$-manifold $(X_M,g_0)$ as in 
Theorem \ref{riemannverzio} and let $S\subset X_M$ be any smoothly embedded, 
connected and orientable (with induced orientation) 
sub-$3$-manifold in it such that with the restricted Riemannian metric 
$h:=g_0\vert_S$ is complete i.e., $(S,h)\subset (X_M,g_0)$ is a 
complete Riemannian sub-$3$-manifold. Of course any compact $S\subset X_M$ 
works but $S$ can be non-compact, too.

\begin{remark}\rm Complete examples $(S,h)\subset (X_M,g_0)$ 
such that $S\subset X_M$ is non-compact can be constructed 
if $S\subset R^4\subset\C P^2$ i.e., it fully belongs to the exotic 
$\R^4$-factor in the decomposition (\ref{ujsokasag}) of $X_M$ as follows. 
The boundary of the unit disk bundle inside the total space of the tautological 
line bundle $H$ over $\C P^1$ is a circle bundle over its zero section 
$\C P^1$ more precisely a Hopf fibration; hence it is a $3$-manifold 
homeomorphic to $S^3$. Fixing an ideal point $x\in\C P^2\setminus R^4$ we can 
identify the total space $H$ with $\C P^2\setminus\{x\}$ and denote by 
$N\subset\C P^2\setminus\{x\}$ the image of the aforementioned boundary of 
the unit disk bundle. Define 
\[S:=\mbox{one connected component of $N\cap R^4$}\:\:\:.\] 
Every exotic $\R^4$ in general hence our $R^4$ in particular, has the property 
that it contains a compact subset $C\subset R^4$ which cannot be surrounded 
by a smoothly embedded $S^3\subset R^4$ \cite[Exercise 9.4.1]{gom-sti}. 
Taking the radii of the constituent circles of $N$ sufficiently 
large we can suppose by the compactness of $C$ that $C\cap S=\emptyset$ i.e., 
$S$ could surround $C$ if $S$ was homeomorphic to $S^3$. This would be a 
contradiction hence $S\subset R^4$ is an open (i.e., non-compact without 
boundary) and connected sub-$3$-manifold of $R^4$. Therefore, exploiting the 
contractibility of $R^4$ we conclude that $S$ is an open contractible 
sub-$3$-manifold within $R^4$. Putting $h:=g_0\vert_S$ we therefore 
obtain an open contractible Riemannian sub-$3$-manifold 
$(S,h)\subset (X_M,g_0)$ which is complete by construction, as the 
reader may verify. 
\end{remark}

\begin{lemma} Consider the Riemannian $4$-manifold $(X_M,g_0)$ 
as in Theorem \ref{riemannverzio} and let $(S,h)\subset 
(X_M,g_0)$ be a connected, oriented and complete Riemannian 
sub-$3$-manifold in it. 

Then there exists a real line sub-bundle $L\subset TX_M$ of the tangent 
bundle such that there exists a smooth Whitney-sum 
decomposition $TX_M=L\oplus L^\perp$ with 
the property that the orthogonal complement (with respect to $g_0$) 
bundle $L^\perp\subset TX_M$ satisfies $L^\perp\vert_S\cong TS$ i.e., 
its restriction is isomorphic to the tangent bundle of $S$.
\label{hasitolemma}
\end{lemma}

\begin{remark}\rm Before we embark upon the proof we clarify that the 
existence of the smooth Whitney-sum decomposition 
$TX_M\cong L\oplus L^\perp$ of the {\it tangent bundle of $X_M$} should 
not be confused with any smooth splitting $X_M\cong \R\times S$ of {\it $X_M$ 
itself}. Indeed, this latter splitting was excluded already in Lemma 
\ref{kezdolemma} once and for all. In fact this non-splitting of 
$X_M$ is a key property of these spaces and is the reason we use 
them throughout the paper.
\end{remark}

\noindent {\it Proof of Lemma \ref{hasitolemma}}. Good references here 
are \cite{hus,ste}. For an oriented Riemannian $4$-manifold standard 
obstruction theory says that the obstruction characteristic classes against 
its tangent bundle being trivial live in the cohomology groups 
$H^i(X_M;\pi_{i-1}({\rm SO}(4)))$, $i=1,\dots, 4$. We know that 
$\pi_0({\rm SO}(4))\cong 0$, 
$\pi_1({\rm SO}(4))\cong\Z_2$, $\pi_2({\rm SO}(4))\cong 0$ and 
$\pi_3({\rm SO}(4))\cong\Z$ but $X_M$ is open 
and oriented hence $H^4(X_M;\Z )\cong 0$. Hence the only obstruction is 
\[w_2(X_M)\in H^2(X_M;\pi_1({\rm SO}(4)))\cong H^2(X_M;\Z_2)\:\:\:,\] 
the so-called $2^{\rm nd}$ Stiefel--Whitney class of $X_M$. 
Consequently if $X_M$ is a spin manifold which by definition means that 
$w_2(X_M)=0$ then its tangent bundle is already trivial hence admits a 
nowhere vanishing smooth section i.e., a non-zero vector field 
$v:X_M\rightarrow TX_M$. Assume $X_M$ is not spin therefore having 
non-trivial tangent bundle. Then exploiting its simply connectivity and 
openness, $X_M$ is homotopic to its $2$-skeleton $X_M(2)$ hence isomorphism 
classes of vector bundles over $X_M$ are in one-to-one correspondence with 
those over $X_M(2)$. However $X_M(2)$ as a topological space is a $2$ 
dimensional CW-complex therefore any 
real rank-$4$ topological vector bundle $E$ over it splits, more 
precisely is isomorphic to $F\oplus\underline{\R}^2$ where $F$ is a real 
rank-$2$ vector bundle and $\underline{\R}^2$ is the trivial real rank-$2$ 
vector bundle. Consequently the tangent bundle $TX_M$ also splits. This of 
course again means that $TX_M$ admits a nowhere vanishing smooth section 
$v:X_M\rightarrow TX_M$ (in fact $TX_M$ admits at 
least two linearly independent sections). 

We construct a section as follows. Taking into account that $S\subset X_M$ 
is orientable, its normal bundle is trivial which means that a small tubular 
neighbourhood $N_\varepsilon (S)\subset X_M$ of $S$ diffeomorphic to 
$S\times (-\varepsilon ,\varepsilon )$. This induces a splitting 
$TX_M\vert_S\cong TS\oplus\underline{\R}$ of the restricted tangent bundle. 
Without loosing generality we can assume that this local splitting is 
orthogonal for the Riemannian metric $g_0$.  Let $v_S:S\rightarrow 
\underline{\R}$ be a nowhere vanishing section of this local orthogonal line 
bundle i.e., $v_S\not=0$ but $g_0(v_S\:,\:TS)=0$. Obstruction theory 
says that $v_S$ can be extended continuously to a section 
$v:X_M\rightarrow TX_M$. Of course this extension is not unique and we can 
arrange it to be smooth and nowhere vanishing because the only obstruction 
class against this latter requirement lives in 
$H^4(X_M;\pi_3(\R^4\setminus\{0\}))
\cong H^4(X_M;\Z )\cong 0$ hence is trivial. The image of this nowhere 
vanishing smooth section $v$ within $TX_M$ then gives rise to a line bundle 
$L\subset TX_M$ and an orthogonal splitting $L\oplus L^\perp =TX_M$ with 
respect to $g_0$ over the whole $X_M$. This splitting satisfies 
$L^\perp\vert_S\cong TS$ by construction, as claimed. $\Diamond$ 
\vspace{0.1in}

\begin{lemma} Take the Ricci-flat Riemannian $4$-manifold $(X_M,g_0)$ 
as in Theorem \ref{riemannverzio} and let $(S,h)\subset (X_M,g_0)$ be a 
connected, oriented and complete Riemannian sub-$3$-manifold.

There exists a smooth Lorentzian metric $g$ on $X_M$ such 
that $(X_M,g)$ is a Ricci-flat Lorentzian manifold (probably null and-or 
timelike incomplete) and $(S,h)\subset (X_M,g)$ is a connected, 
complete spacelike sub-$3$-manifold.
\label{lorentzlemma}
\end{lemma}

\begin{remark}\rm We emphasize that $g$ is not the result of an analytic
continuation of $g_0$ within some complex manifold hence the procedure
described in Lemmata \ref{hasitolemma} and \ref{lorentzlemma} is not a
``Wick rotation'' in any sense of e.g. \cite{hel-her} and the references
therein. Accordingly, $g$ is not uniquely determined by $g_0$, it depends on
the chosen subspace $S\subset X_M$ and more generally, the line bundle
$L\subset TX_M$. The main reason for not using the standard approach,
beyond its rigidity, is that we do not want to loose the subtle smoothness
properties of $X_M$ by replacing it with another manifold within its
complexification $X^\C_M$ which, by twistor theory, exists
(cf. the {\it Remark} on twistor theory in Section \ref{three}).
\end{remark}

\noindent {\it Proof of Lemma \ref{lorentzlemma}}. Take the 
complexification $T^\C X_M:=TX_M\otimes_\R\C$ 
of the real tangent bundle as well as the complex 
linear extension of the Riemannian Ricci-flat metric $g_0$ on $TX_M$ to a 
complex Ricci-flat metric $g^\C_0$ on $T^\C X_M$. This means that if 
$v^\C$ is a complex tangent vector then both $v^\C\mapsto 
g^\C_0(v^\C\:,\:\cdot\:)$ and $v^\C\mapsto g^\C_0(\:\cdot\:,\:v^\C)$ 
are $\C$-linear maps and ${\rm Ric}^\C=0$. Then the real 
splitting $TX_M=L\oplus L^\perp$ of Lemma \ref{hasitolemma}, satisfying 
$L^\perp\vert_S\cong TS$ with the chosen $(S,h)\subset (X_M,g_0)$, induces a 
splitting 
\begin{equation}
T^\C X_M=L\oplus L^\perp\oplus\sqrt{-1}\:L\oplus\sqrt{-1}\:L^\perp
\label{felbontas}
\end{equation}
over $\R$ i.e., if $T^\C X_M$ considered as a real 
rank-$8$ bundle over $X_M$. Define a metric on the real rank-$4$ sub-bundle 
$L^\perp\oplus\sqrt{-1}\:L\subset T^\C X_M$ by taking the restriction 
$g_0^\C\vert_{L^\perp\oplus\sqrt{-1}\:L}$. It readily follows from the 
orthogonality of the splitting that this is a non-degenerate real bilinear 
form of Lorentzian type on this real sub-bundle. To see this, we simply have 
to observe that with real vector fields $v_1,v_2:X_M\rightarrow L$ 
and $w_1,w_2:X_M\rightarrow L^\perp$  
\[g_0^\C\vert_{L^\perp\oplus\sqrt{-1}\:L}(\sqrt{-1}\:v_1,\sqrt{-1}\:v_1)= 
g_0^\C(\sqrt{-1}\:v_1,\sqrt{-1}\:v_1)=-g^\C_0(v_1,v_1) =-g_0(v_1,v_1)\]  
and 
\[g_0^\C\vert_{L^\perp\oplus\sqrt{-1}\:L}(\sqrt{-1}\:v_1,w_1)=
g_0^\C(\sqrt{-1}\:v_1,w_1)=\sqrt{-1}\:g^\C_0(v_1,w_1) 
=\sqrt{-1}g_0(v_1,w_1)=0\]
and finally 
\[g_0^\C\vert_{L^\perp\oplus\sqrt{-1}\:L}(w_1,w_2)=
g_0^\C(w_1,w_2)=g_0(w_1,w_2)\:\:\:.\]
Consider the $\R$-linear bundle isomorphism $W_L:T^\C X_M\rightarrow T^\C 
X_M$ of the complexified tangent bundle defined by, with 
respect to the splitting (\ref{felbontas}), as 
$W_L(v_1,w_1,\sqrt{-1}\:v_2,\sqrt{-1}\:w_2):=
(v_2, w_1,\sqrt{-1}\:v_1,\sqrt{-1}\:w_2)$. It maps the real tangent bundle 
$TX_M=L\oplus L^\perp\subset T^\C X_M$ onto the real bundle 
$L^\perp\oplus\sqrt{-1}\:L\subset T^\C X_M$ and {\it vice versa} 
making the diagram
\[\xymatrix{
\ar[d] T^\C X_M\ar[r]^{W_L} & T^\C X_M\ar[d] \\
          X_M\ar[r]^{{\rm Id}_{X_M}}  & X_M   
}\]
commutative. In fact $W_L$ is a real {\it reflection} satisfying 
$W_L^2={\rm Id}_{T^\C X_M}$. Then with arbitrary two tangent vectors 
$v,w:X_M\rightarrow TX_M$ putting 
\[g(v,w):=g_0^\C(W_Lv\:,\:W_Lw)\] 
we obtain a Lorentzian metric $g$ on $TX_M$ such that $(S,h)\subset 
(X_M,g)$ is a connected, complete spacelike sub-$3$-manifold. 

Concerning its Ricci tensor, the Levi--Civita 
connection $\nabla$ of $g$ and $\nabla^\C$ of $g^\C_0$ are related by 
\begin{eqnarray}
g(\nabla_uv,w)+g(v,\nabla_uw)&=&\dd g(v,w)u =\dd 
g^\C_0(W_Lv,W_Lw)u\nonumber\\
&=&g^\C_0(\nabla^\C_u(W_Lv),W_Lw)+g^\C_0(W_Lv, 
\nabla^\C_u(W_Lw))\nonumber
\end{eqnarray}
yielding $\nabla=W_L\nabla^\C W_L$ (as an $\R$-linear 
operator) consequently the curvature ${\rm Riem}$ of $g$ takes the shape  
\[{\rm Riem}(v,w)u=\left[\nabla_v\:,\nabla_w\right]u-
\nabla_{[v,w]}u=W_L({\rm Riem}^\C(v,w)W_Lu)\:\:.\]
Let $\{e_0,e_1,e_2,e_3\}$ be a real orthonormal frame for $g$ at $T_pX_M$ 
satisfying $g(e_0,e_0)=-1$ and $+1$ for the rest, then 
$W_Le_0=\sqrt{-1}\:e_0$ and $W_Le_j=e_j$ for $j=1,2,3$ together 
with the definition of $g$ imply first that 
\[g({\rm Riem}(e_0,v)w,e_0)=g_0^\C\left(W_L({\rm 
Riem}(e_0,v)w),W_Le_0\right)=g_0^\C\left({\rm 
Riem}^\C(e_0,v)W_Lw\:,\sqrt{-1}e_0\right)\]
and likewise  
\[g({\rm Riem}(e_j,v)w,e_j)=g_0^\C(W_L({\rm 
Riem}(e_j,v)w),W_Le_j)=g_0^\C({\rm 
Riem}^\C\left(e_j\:,v\right)W_Lw\:,e_j)\:\:.\]
The Ricci tensor in any signature looks like 
${\rm Ric}(v,w)=\sum_{k=1}^mg(e_k,e_k)g({\rm Riem}(e_k,v)w,e_k)$; hence
\begin{eqnarray}
{\rm Ric}(v,w)\!\!\!\!\!\!&=&\!\!\!\!\!\!g(e_0,e_0)g({\rm 
Riem}(e_0,v)w,e_0)+
\sum\limits_{j=1}^3g(e_j,e_j)g({\rm Riem}(e_j,v)w,e_j)\nonumber\\
\!\!\!\!\!\!&=&\!\!\!\!\!\!g^\C_0(\sqrt{-1}e_0,\!\sqrt{-1}e_0)g^\C_0({\rm 
Riem}^\C(e_0,v)W_Lw,\sqrt{-1}e_0)
\!+\!\!\sum\limits_{j=1}^3g^\C_0(e_j,e_j)g^\C_0({\rm 
Riem}^\C(\!e_j,v)W_Lw,e_j\!)\nonumber\\
\!\!\!&=&\!\!\!-(\sqrt{-1}\:+1)g^\C_0(e_0,e_0)g^\C_0({\rm 
Riem}^\C(e_0,v)W_Lw,e_0)+{\rm Ric}^\C(v\:,\:W_Lw)\nonumber\\
&=& (\sqrt{-1}-1)g({\rm Riem}(e_0,v)w,e_0)\nonumber
\end{eqnarray}
and we also used $\{e_0,e_1,e_2,e_3\}$ as a complex basis to obtain 
$\sum_{j=0}^3g^\C_0(e_j,e_j)g^\C_0({\rm Riem}^\C(e_j,v)W_Lw,e_j)=
{\rm Ric}^\C(v,W_Lw)=0$. However the last expression can be real for all 
$v,w\in T_pX_M$ if and only if it vanishes. This demonstrates that $g$ is 
indeed Ricci-flat. $\Diamond$
\vspace{0.1in}

\noindent After this very long technical journey through Sections \ref{three} 
and \ref{four} we are now ready to inspect $(X_M,g)$ concerning its global 
hyperbolicity. 

\begin{lemma} Consider the Ricci-flat Lorentzian $4$-manifold $(X_M,g)$ 
of Lemma \ref{lorentzlemma} with any spacelike and 
complete sub-$3$-manifold $(S,h)\subset (X_M,g)$ in it (Lemma 
\ref{lorentzlemma} also provides us that non-empty submanifolds of 
this sort exist). Let $(S,h,k)$ be the initial data 
set inside $(X_M,g)$ induced by $(S,h)$ and let $(X'_M,g')$ be a 
perturbation of $(X_M,g)$ relative to $(S,h,k)$ as in Definition 
\ref{perturbacio}. Consider the pair $(R^4,  K)$ from Theorem 
\ref{egzotikusnagycsalad}. Assume that $X_M'$ contains the 
image, present in the $R^4$-factor of $X_M$ in its decmoposition 
(\ref{ujsokasag}), of the compact subset $K$. Then $(X'_M,g')$ is not 
globally hyperbolic. 
\label{ellenpeldalemma}
\end{lemma}

\noindent {\it Proof.} First we prove that the 
trivial perturbation i.e., $(X_M, g)$ itself is not globally hyperbolic. To 
see this observe that $X_M$ is not a product of any $3$-manifold $S$ and $\R$ 
due to its creased end (cf. Lemma \ref{kezdolemma}); hence it follows from the 
smooth splitting theorem for globally hyperbolic space-times \cite{ber-san} 
that $(X_M,g)$ cannot be globally hyperbolic.

Let us secondly consider its non-trivial perturbations $(X'_M,g')$ 
relative to $(S,h,k)$ as in Definition \ref{perturbacio}. Suppose that 
$(X'_M,g')$ is globally hyperbolic. Referring to Definition \ref{perturbacio} 
we know that $(S,h')\subset (X_M',g')$ is a complete spacelike submanifold 
hence we can use it to obtain an initial data set $(S,h',k')$ for $(X_M',g')$. 
Again by \cite{ber-san} we find $X_M'\cong S\times\R$. 
But by our Definition \ref{perturbacio} the perturbed space always satisfies 
$S\subset X'_M\subseteqq X_M$ and, by our assumption in the present lemma, 
$X'_M$ still contains the image of the compact subset $K\subset R^4$. 
This means that there exists a connected smooth $4$-manifold $M'$ satisfying 
\[S\subset M'\subseteqq M\#\underbrace{\C P^2\#\dots\#\C P^2}_{k-1}\] 
and an exotic $R^4_t$ with $0<r\leqq t\leqq +\infty$ from the family in part 
(ii) of Theorem \ref{egzotikusnagycsalad} such that $X'_M$ has a 
decomposition $X'_M\cong M'\#_KR^4_t$, too. 
However this is in a contradiction with the splitting of $X'_M$ above. 
This demonstrates that our supposition was wrong hence $(X_M',g')$ cannot be 
globally hyperbolic. $\Diamond$
\vspace{0.1in}

\begin{remark}\rm The condition that the perturbed space $X'_M$ 
should contain the compact subset $K\subset R^4$ can be interpreted as follows. 
Decomposition (\ref{ujsokasag}) shows that $X_M$ has only one asymptotic 
region namely its creased end from its exotic $\R^4$-component (see Figure 1). 
Therefore the Lorentzian manifold $(X_M,g)$ can be regarded as a vacuum 
space-time describing some topologically non-trivial ``inner'' region 
corresponding to $M\#\C P^2\#\dots\#\C P^2$ and a contractible surrounding 
``outer'' region described by $R^4$ in the decomposition (\ref{ujsokasag}) of 
$X_M$. The condition that the perturbation satisfying 
$S\subset X'_M\subseteqq X_M$ should contain the compact subset $K\subset 
R^4$ present in the original space-time $X_M$ means, taking into account 
the precise glueing descriptions in Lemma \ref{kezdolemma}, that $X'_M$ 
yet contains a ``sufficiently large part'' of the original space $X_M$ 
i.e., cannot be simply e.g. a small tubular neighbourhood $S\subset 
N_\varepsilon (S)\subset X_M$ of the initial surface. Therefore this 
simple assumption says that the perturbation about $S\subset X_M$ is large 
enough in the topological sense hence is capable to ``scan'' the exotic 
regime of $X_M$. More on the physical interpretation of $(X_M,g)$ we refer 
to Section \ref{five}.

In fact this condition is effectively necessary to exclude globally 
hyperbolic perturbations of the original space $(X_M\:,\:g)$. Let $M:=S^4$ 
then $X_{S^4}=R^4$ and let $S\subset X_{S^4}$ be any connected open 
sub-$3$-manifold in it; then putting $X'_{S^4}:=N_\varepsilon (S)\subset 
X_{S^4}$ to be a small tubular neighbourhood of $S\subset X_{S^4}$ the 
contractibility of $S$ implies $N_\varepsilon (S)\cong S\times\R$ hence 
again by \cite{mcm} we know that $N_\varepsilon (S)\cong\R^4$. Therefore 
putting $g'$ just to be the standard Minkowski metric on $X'_{S^4}$ we 
obtain $(X'_{S^4},g')$ is the usual Minkowski space-time hence is a 
globally hyperbolic perturbation of $(X_{S^4},g)$ relative to $(S,h,k)$. 
This perturbation is ``small'' in the topological sense above however 
might be ``large'' in any analytical sense i.e., the corresponding 
$(S,h',k')$ might siginificantly deviate from the original $(S,h,k)$. 
\end{remark}

\noindent {\it Proof of Theorem \ref{lorentzverzio}}. Putting together 
Lemmata \ref{hasitolemma}, \ref{lorentzlemma} and \ref{ellenpeldalemma} 
we obtain the result. $\Diamond$
\vspace{0.1in}

\noindent Finally we are in a position to draw the main conclusion 
of our efforts so far, namely to put the immense class of vacuum 
space-times we found in the context of the {\bf SCCC}. Taking into account 
the non-trivial condition regarding 
$K\subset R^4$ in Lemma \ref{ellenpeldalemma}, the space $(X_M,g)$ is not a 
robust counterexample to the {\bf SCCC} in the strict sense of Definition 
\ref{ellenpelda}. However knowing that we can start with {\it any} 
closed and simply connected $M$ to construct open spaces like $X_M$ with a 
creased end carrying a solution $g$ of the Lorentzian vacuum Einstein 
equation, and the class of non-globally-hyperbolic perturbations 
$(X'_M,g')$ of $(X_M,g)$ are subject only to this mild topological condition, 
the corresponding perturbation class is certainly still enormously vast. 
Therefore in our opinion it is reasonable to say that {\it all the members of 
these immense family of Lorentzian vacuum space-times $(X_M,g)$ give 
rise to generic counterexamples to the {\bf SCCC}} as formulated in the 
Introduction (recall that being generic is not a well-defined concept). 
This is the content of the informal statement $\overline{\bf SCCC}$, also 
formulated in the Introduction. In other words, in light of our 
consderations so far: {\it the {\bf SCCC} typically fails in four dimensions!}


\section{Conclusion and outlook}
\label{five}


From the viewpoint of low dimensional differential topology it is not 
surprising that confining ourselves into the initial value approach when 
thinking about the {\bf SCCC} typically brings affirmative while more 
global techniques might yield negative answers: the initial 
value formulation of Einstein's equation likely just explores 
the vicinity of $3$ dimensional smooth spacelike 
submanifolds inside the full $4$ dimensional space-time. It is 
well-known that an embedded smooth submanifold of an ambient space 
always admits a tubular neighbourhood which is an open disk bundle over 
the submanifold i.e., has a locally product smooth structure. However exotic 
$4$ dimensional smooth structures never arise as products of lower 
dimensional ones consequently the {\it four} dimensional exotica i.e., 
the general structure of space-time never can be detected from a {\it three} 
dimensional perspective such as the initial value formulation. 
There is a {\it qualitative leap} between these dimensions. 

The physical interpretation of the vacuum solutions $(X_M,g)$ 
is a challenging question because, as we 
have seen in the {\it Remark} after Lemma \ref{teljeslemma} the 
corresponding Riemannian spaces $(X_M,g_0)$ are all gravitational 
instantons hence are dominantly present if some quantum theory lurks behind; 
i.e., although they might play no role in classical general 
relativity, the interpretation of these solutions is unavoidable in a 
broader quantum context. As we already observed in the {\it 
Remark} after Lemma \ref{ellenpeldalemma}, $(X_M, g)$ is a vaccum 
space-time such that $X_M$ is simply connected consisting of a topologically 
non-trivial interior part and a topologically trivial asymptotic region; 
however the metric $g$ on this space surely cannot decay 
to the flat metric because this asymptotic region is creased and $g$ 
still has a Weyl tensor. Consequently $(X_M,g)$ is not the 
geometry of a ``compact gravitating system'' or anything like that. On the 
contrary, its peculiarity is its asymptotical structure on its creased 
end. Exotic phenomena are genuinely non-local in the sense that all 
$4$-manifolds are locally the same therefore, in our understanding at the 
current state of art, these vacuum solutions with their distant creased 
properties rather correspond to (quantum)cosmological solutions. The inherent 
non-global-hyperbolicity of them very likely stems from the violation of strong 
causality along their creased end probably caused by the fractal-like 
behaviour of distant spacelike submanifolds. Indeed, we already mentioned 
in the {\it Remark} before Lemma \ref{hasitolemma} that exotic $\R^4$'s 
have the property that sufficiently large compact subsets of them cannot 
be surrounded by smoothly embedded $S^3$'s. The 
non-deterministic character of these ``(quantum)cosmological solutions'' 
towards their infinity could perhaps be physically understood as the 
manifestation of the quantum properties of the Big Bang singularity.


\section{Appendix: Lebesgue integration in algebraic function fields}
\label{six}


Here we work out how to integrate functions on manifolds but taking values 
in the algebraic function field of 
complex rational functions. The construction of this integral is 
straightforward and is fully based on the by-now classical approach of 
Riesz--Sz\H okefalvi-Nagy (cf. \cite[\S 16, \S 17]{rie-szo}).

Let $P(z),Q(z)$ be two complex polynomials in the single variable 
$z\in\C$ and let $\C (z)$ denote the (commutative) algebraic function field 
of fractions $R(z):=\frac{P(z)}{Q(z)}$. A norm of $R$ is defined by the formula 
$\vert R\vert_{c,z_0}:=c^{{\rm ord}_{z_0}(R)}$ where $c\in (0,1)$ is a fixed 
real number and ${\rm ord}_{z_0}(R)\in\Z$ is the lowest one among the indicies 
$k\in\Z$ of the non-zero coefficients $a_k\in\C$ in the Lauent expansion 
\begin{equation}
R(z)=\sum\limits_{k\gg -\infty}^{+\infty}a_k(z-z_0)^k
\label{laurent}
\end{equation}
of $R$ about a fixed point $z_0\in\C$; note that the number 
${\rm ord}_{z_0}(R)$ is independent of the particular coordinate system 
used for the expansion hence is well-defined and this definition makes 
sense for $R=0$ if we put ${\rm ord}_{z_0}(0):={+\infty}$ and of course 
yields $\vert 0\vert_{c,z_0}=0$. It is known \cite[Theorem 1.11]{stev} 
that, being $\C$ algebraically closed, $\vert\:\cdot\:\vert_{c,z_0}$ with 
$c\in (0,1)$ and $z_0\in\C\cup\{\infty\}$ is the complete list of norms on 
$\C (z)$ which are trivial on $\C$. Then $\C (z)$ can be completed with respect 
to $\vert\:\cdot\:\vert_{c,z_0}$ which is $\C (\!(z-z_0)\!)$, the field of 
formal Laurent series in $z-z_0$. There is an embedding of fields 
$\C (z)\subset\C (\!(z-z_0)\!)$ for all $c,z_0$ but up to 
isomorphisms of topological fields, these completions are independent 
of the norms used \cite{stev}. 

\begin{remark}\rm Unlike the usual norms on $\R$ or $\C$, all the norms of 
this kind on $\C (z)$ are {\it non-Archimedean} hence $\C (z)$ does not 
admit any norm-compatible embedding into $\C$ consequently its 
analytical properties are quite different from those of the real or 
complex numbers. Moreover the spectra of our norms here are {\it 
discrete}, more precisely $\vert\C (\!(z-z_0)\!)\vert_{c,z_0}=c^\Z\subset 
[0,+\infty]$ consequently the spectrum of $\vert\:\cdot\:\vert_{c,z_0}$ 
for all $c,z_0$ has only one accumulation point $0\in\R$. Another essential 
difference is that, unlike $\R$, the algebraic function field $\C (z)$ is 
{\it not ordered}.
\end{remark}

\noindent Let $(M,g)$ be an oriented Riemannian $m$-manifold. Then $M$ is 
equipped with a measure ${\rm vol}_g$ coming from the volume-form 
$\dd{\rm vol}_g:=*_g1$ provided by the orientation and the metric; the 
corresponding measure of a measurable subset 
$\emptyset\subseteqq A\subseteqq M$ is 
${\rm vol}_g(A):=\int_M\chi_A\dd\:{\rm vol}_g=\int_A\dd\:{\rm vol}_g$ where 
$\chi_B:M\rightarrow\{0,1\}$ is the characteristic function of any subset 
$\emptyset\subseteqq B\subseteqq M$. Clearly $0\leqq{\rm vol}_g(A)\leqq 
+\infty$ is a non-negative (extended) real number. Take finitely many 
measurable subsets $U_1,\dots, U_n\subset M$ whose closures are coordinate 
balls of finite volume but are pairwise almost non-intersecting; that 
is, $U_i\subset M$ has the property that $\overline{U}_i$ is 
diffeomorphic to the standard closed ball $\overline{B}^m\subset\R^m$ 
moreover $0<{\rm vol}_g(U_i)<+\infty$ for all $i$ but 
${\rm vol}_g(U_i\cap U_j)=0$ for all $i\not=j$. Also take elements 
$R_1,\dots,R_n\in\C (\!(z-z_0)\!)$. A function 
$\varphi : M\rightarrow\C (\!(z-z_0)\!)$ of the form 
\[\varphi :=\sum\limits_{j=1}^nR_j\:\chi_{U_j}\]
is called an {\it elementary step function}. 
This definition makes sense since $\R$ acts on $\C (\!(z-z_0)\!)$ by 
multiplication; nevertheless these functions might be ill-defined in 
boundary points however, as we already anticipate from Lebesgue theory, 
ambiguities of this sort will be negligable concerning their integrals. The 
{\it integral} of an elementary step function against the measure induced 
by $\dd\:{\rm vol}_g$ is defined as  
\[\int\limits_M\varphi\:\dd{\rm vol}_g:=\sum\limits_{j=1}^nR_j\:{\rm vol}_g(U_j)\in
\C (\!(z-z_0)\!)\] 
($\varphi$ is written sometimes as $R: M\rightarrow \C (\!(z-z_0)\!)$ and its 
integral as $\int_M R_x\:\dd x$\:, too) in full analogy with the usual case. 

Next let us recall the two elementary but fundamental lemmata from 
\cite{rie-szo} what we state here in appropriately modified forms and 
prove as follows. 

\begin{lemma} {\rm (cf. \cite[Lemme {\bf A}, p. 30]{rie-szo})} Let 
$\{\varphi_i\}_{i\in\N}$ be a sequence of elementary step functions from a 
compact oriented Riemannian manifold $(M,g)$ into $\C (\!(z-z_0)\!)$. If 
$\{\vert\varphi_i\vert_{c,z_0}\}_{i\in\N}$ is strictly decreasing almost 
everywhere, then the integrals of these functions converge to zero in 
$\C (\!(z-z_0)\!)$. 
\label{lemmaa}
\end{lemma} 

\noindent {\it Proof.} As mentioned above the spectrum of 
$\vert\:\cdot\:\vert_{c,z_0}$ has only one limit point $0\in\R$ 
therefore if $\{\vert\varphi_i\vert_{c,z_0}\}_{i\in\N}$ {\it strictly} 
decreases almost everywhere then in fact 
$\vert\varphi_i(x)\vert_{c,z_0}\rightarrow 0$ if 
$x\in M\setminus B$ as $i\rightarrow +\infty$ where $B$ is a subset of 
measure zero i.e., for any $\delta\geqq 0$ there exist open subsets 
$V_\delta\subset M$ satisfying ${\rm vol}_g(V_\delta)\leqq\delta$ such 
that $\emptyset\subseteqq B\subset V_\delta$. This means 
on the one hand that for every $\varepsilon\geqq 0$ there exists an index 
$i_\varepsilon$ such that for all $i\geqq i_\varepsilon$ one finds
\[0\leqq\left\vert\:\:\int\limits_{M\setminus B}\varphi_i\:\dd{\rm vol}_g
\right\vert_{c,z_0}\leqq
\sup\limits_{x\in M\setminus B}\vert\varphi_i (x)\vert_{c,z_0}
\leqq\varepsilon\:\:\:.\]
On the other hand, if for any fixed $i$ and $x\in B$ the lowest non-zero 
coefficient of $\varphi_i(x)$ in (\ref{laurent}) is 
$a_{iK}(x)\in\C$ then the same coefficient of $\int_B\varphi_i\:
\dd{\rm vol}_g$ can be estimated from above by 
\[\sup_{x\in B}\vert a_{iK}(x)\vert{\rm vol}_g(V_\delta)
\leqq\sup_{x\in B}\vert a_{iK}(x)\vert\delta\] 
and exploiting the compactness of $M$ we can assume that the number of 
the different leading coefficients $a_{iK}(x)$ is finite as $x$ runs over 
$B$ hence surely $\sup_{x\in B}\vert a_{iK}(x)\vert <+\infty$. It then follows 
that the leading coefficient of the integral is arbitrary small hence 
$\vert\:\int_B\varphi_i\:\dd{\rm vol}_g\vert_{c,z_0}=0$ i.e., the integral 
over $B$ vanishes for every fixed index $i$. Consequently 
$0\leqq\vert\:\int_M\varphi_i\:\dd{\rm vol}_g\vert_{c,z_0}\leqq\varepsilon$ 
for all $i\geqq i_\varepsilon$. That is, the sequence of integrals converges 
to zero as stated. $\Diamond$ 
\vspace{0.1in}

\begin{lemma} {\rm (cf. \cite[Lemme {\bf B}, p. 30]{rie-szo})}  Let
$\{\varphi_i\}_{i\in\N}$ be a sequence of elementary step functions from an 
oriented Riemannian manifold $(M,g)$ into $\C (\!(z-z_0)\!)$. If 
$\{\vert\varphi_i\vert_{c,z_0}\}_{i\in\N}$ is increasing and the sequence 
$\{\int_M\varphi_i\:\dd{\rm vol}_g\}_{i\in\N}$ of the corresponding integrals 
converges to an element in $\C (\!(z-z_0)\!)$, then $\{\varphi_i\}_{i\in\N}$ 
converges to a finite limit in $\C (\!(z-z_0)\!)$ almost everywhere. 
\label{lemmab}
\end{lemma}

\noindent {\it Proof.} Let $\emptyset\subseteqq B\subseteqq M$ denote the 
collection of all of those points $x\in M$ where 
$\varphi_i(x)$ is divergent in $\C (\!(z-z_0)\!)$ as $i\rightarrow+\infty$. 
This can mean two (not necessarily mutually exclusive) 
things: either a sequence $\{a_{ik_i}(x)\}_{i\in\N}$ of coefficients in 
the expansions (\ref{laurent}) of the $\varphi_i(x)$'s is divergent or 
$\{\vert\varphi_i(x)\vert_{c,z_0}\}_{i\in\N}$ is divergent i.e., the index 
set $\{K_i\}_{i\in\N}$ of the lowest non-zero $a_{iK_i}(x)$'s in the expansions 
of the $\varphi_i(x)$'s with $x\in B$ is unbounded from below. In either cases, 
since the sequence of the corresponding integrals 
$\int_M\varphi_i\:\dd{\rm vol}_g$ converges to a well-defined element 
$R\in\C (\!(z-z_0)\!)$ with a well-defined expansion (\ref{laurent}) whose 
coefficients are of the form $a_k{\rm vol}_g(U)$, these divergences can be 
absent from the integral if and only if for every $\delta\geqq 0$ there exist 
open subsets $\emptyset\subseteqq B\subset V_\delta\subset M$ such that 
${\rm vol}_g(V_\delta )\leqq\delta$. 
In other words $B$ is of measure zero as stated. $\Diamond$   
\vspace{0.1in}

\noindent From here we proceed in the standard way 
(cf. \cite[\S 17]{rie-szo}) hence we only quickly summarize the main 
steps. Let $(M,g)$ be a compact oriented Riemannian manifold. 
If $C_0(M;\C (\!(z-z_0)\!))$ is the set of elementary step functions from 
$M$ to $\C (\!(z-z_0)\!)$ then let $C_1(M;\C (\!(z-z_0)\!))$ denote the 
set of those functions $f:M\rightarrow \C (\!(z-z_0)\!)$ which arise as 
limits of sequences of functions $\{\varphi_i\}_{i\in\N}$ in 
Lemma \ref{lemmab} i.e., arise almost everywhere as the limits 
$f(x):=\lim\varphi_i(x)$ of increasing elementary step functions with a 
convergent sequence of corresponding integrals. Define their {\it 
integrals}, which therefore exist, to be
\[\int\limits_Mf\:\dd{\rm vol}_g:=\lim\limits_{i\rightarrow +\infty}
\int\limits_M\varphi_i\:\dd{\rm vol}_g\in\C (\!(z-z_0)\!)\] 
(again $f$ is written sometimes as $R: M\rightarrow \C 
(\!(z-z_0)\!)$ and its integral as $\int_M R_x\:\dd x$\:, too). 
This definition is correct because, by referring to 
Lemma \ref{lemmaa}, the integral does not depend on the particular choice of 
the sequence $\{\varphi_i\}_{i\in\N}$ converging almost everywhere to a 
given $f$. The set $C_1(M;\C (\!(z-z_0)\!))$ has already the structure of a 
vector space over $\C (\!(z-z_0)\!)$ and is closed and complete in an 
appropriate sense; it is more commonly denoted as $L^1(M;\C (\!(z-z_0)\!))$ 
and called the {\it space of $\C (\!(z-z_0)\!)$-valued Lebesgue integrable 
functions on $M$} (with respect to a measure coming from the orientation 
and metric on $M$). 

The main purpose of these investigations is to complete the proof of 
Lemma \ref{pilemma} by demonstrating 
\begin{lemma} 
Using the notations of Lemma \ref{pilemma}, the map $\pi: 
Z(R^4)\rightarrow p^{-1}([{\mathfrak l}_0])$ constructed 
by the integral (\ref{azintegral}) is well-defined and holomorphic. 
Consequently for every fixed $x\in\C P^2\setminus R^4$ this integral 
satisfies $\int_{S^2}R_{i^*y,x}(\pi_x(\divideontimes))\dd (i^*y)\in\C (\pi_x)$ 
i.e., is again a rational function in the variable $\pi_x$. 

Moreover, picking two points $x_1, x_2\in \C P^2\setminus R^4$ and for 
every $({\mathfrak l},{\mathfrak p})\in Z(R^4)$ we find 
\[\int\limits_{S^2}R_{i^*y,x_1}(\pi_{x_1}(({\mathfrak l},{\mathfrak p})))
\dd (i^*y)=
\int\limits_{S^2}R_{i^*y,x_2}(\pi_{x_2}(({\mathfrak l},{\mathfrak p})))
\dd (i^*y)\]
hence taking into account the relation 
\[\pi_{x_2}(({\mathfrak l},{\mathfrak p}))=\frac{P_{x_2,x_1}(\pi_{x_1}
(({\mathfrak l},{\mathfrak p})))}{Q_{x_2,x_1}
(\pi_{x_1}(({\mathfrak l},{\mathfrak p})))}\] 
as well, the change of the reference point $x$ in (\ref{azintegral}) can 
be regarded as an algebraic change of variables in the 
coordinatized algebraic function field of rational functions. Therefore 
one can talk about an integral $\int_{S^2}R_{i^*y}\:\dd (i^*y)\in\C (z)$ 
in an abstract sense. 
\label{holomorflemma}
\end{lemma}

\noindent {\it Proof.} First of all it easily follows that $\pi$ exists 
since taking any $({\mathfrak l},{\mathfrak p})\in Z(R^4)$ the 
corresponding extended complex number $\pi (({\mathfrak l},
{\mathfrak p}))\in p^{-1}([{\mathfrak l}_0])\cong\C P^1$ is well-defined 
because it arises as the particular value of a Lebesgue 
integral; and this Lebesgue integral exists by the appropriate version of 
Lebesgue's dominated convergence theorem applied to the sequence of 
step functions converging almost everywhere to the bounded function 
$i^*y\mapsto R_{i^*y,x}$ on the compact $S^2$ equipped with its standard 
orientation and metric providing a measure $\dd (i^*y)$ on it. Regarding its 
holomorphicity, observe that the integral in (\ref{azintegral}) is nothing 
else than a limit:  
\[\int\limits_{S^2}R_{i^*y,x}(\pi_x(({\mathfrak l},{\mathfrak p})))\:\dd (i^*y)
=\lim\limits_{n\rightarrow +\infty} 
\sum\limits_{j=1}^nR_{j,x}(\pi_x(({\mathfrak l},{\mathfrak p}))) 
{\rm vol} (U_j)=\lim\limits_{n\rightarrow +\infty}
\sum\limits_{j=1}^n\frac{P_{j,x}(\pi_x(({\mathfrak l},{\mathfrak p})))}
{Q_{j,x}(\pi_x(({\mathfrak l},{\mathfrak p})))}\:{\rm vol} (U_j)\]
of integrals of elementary step functions. All the terms in 
these sums hence the sums themselves for any finite $n$ are holomorphic. 
Holomorphicity here means that if 
\[I_{({\mathfrak l},{\mathfrak p})}:T_{({\mathfrak l},
{\mathfrak p})}Z(R^4)\longrightarrow T_{({\mathfrak l},   
{\mathfrak p})}Z(R^4)\]
is the induced almost complex operator of $Z(R^4)$ at $({\mathfrak l},
{\mathfrak p})$ and 
\[\pi_x(({\mathfrak l},{\mathfrak p}))_*\::\:T_{({\mathfrak l},
{\mathfrak p})}Z(R^4)\longrightarrow 
T_{\pi_x(({\mathfrak l},{\mathfrak p}))}p^{-1}([{\mathfrak l}_0])\] 
is the derivative of $\pi_x:Z(R^4)\rightarrow p^{-1}([{\mathfrak l}_0])$ 
at $({\mathfrak l},{\mathfrak p})$ and 
\[J_{\pi_x(({\mathfrak l},{\mathfrak p}))}:T_{\pi_x(({\mathfrak l},
{\mathfrak p}))}p^{-1}([{\mathfrak l}_0])\longrightarrow 
T_{\pi_x(({\mathfrak l},{\mathfrak p}))}p^{-1}([{\mathfrak l}_0])\] 
is the induced almost complex operator of $p^{-1}([{\mathfrak l}_0])
\cong\C P^1$ at $\pi_x(({\mathfrak l},{\mathfrak p}))$ then the derivatives 
\[\left(R_{j,x}(\pi_x(({\mathfrak l},{\mathfrak p})))
{\rm vol} (U_j)\right)_*\::\: T_{({\mathfrak l},{\mathfrak p})}Z(R^4)
\longrightarrow 
T_{\pi_x(({\mathfrak l},{\mathfrak p}))}p^{-1}([{\mathfrak l}_0])\]
of the individual terms in the sum above, equal to 
\[\left(\frac{P'_{j,x}(\pi_x(({\mathfrak l},{\mathfrak p})))
Q_{j,x}(\pi_x(({\mathfrak l},{\mathfrak p})))-P_{j,x}
(\pi_x(({\mathfrak l},{\mathfrak p})))Q'_{j,x}(\pi_x(({\mathfrak l},
{\mathfrak p})))}{Q^2_{j,x}(\pi_x(({\mathfrak l},{\mathfrak p})))}\:
{\rm vol} (U_j)\right)
\pi_x(({\mathfrak l},{\mathfrak p}))_*\:\:\:,\] 
commute with the almost complex operators i.e., 
\[\left(R_{j,x}(\pi_x(({\mathfrak l},{\mathfrak p})))
{\rm vol} (U_j)\right)_*\circ I_{({\mathfrak l},{\mathfrak p})} 
=J_{\pi_x(({\mathfrak l},{\mathfrak p}))}\circ 
\left(R_{j,x}(\pi_x(({\mathfrak l},{\mathfrak p}))) {\rm vol} 
(U_j)\right)_*\]
for each $j=1,2,\dots,n$. However this property obviously survives the 
limit $n\rightarrow +\infty$ to be taken. 

Last but not least, for every fixed $y\in\C P^2\setminus R^4$ the map 
$\pi_y:Z(R^4)\rightarrow p^{-1}([{\mathfrak l}_0])$ constructed in 
(\ref{pix}) is well-defined consequently
\[R_{y,x_1}(\pi_{x_1}(({\mathfrak l},{\mathfrak p})))=\pi_y(({\mathfrak l},
{\mathfrak p}))=R_{y,x_2}(\pi_{x_2}(({\mathfrak l},{\mathfrak p})))\]
demonstrating the equality of the corresponding integrals. $\Diamond$

\end{document}